\RequirePackage[T1]{fontenc}
\documentclass[a4paper,11pt]{article}
\usepackage{jheppub} 
\usepackage[utf8]{inputenc}

\usepackage{jheppub}
\usepackage[utf8]{inputenc}
\usepackage{amsmath}
\usepackage{amssymb}
\usepackage{mathtools}
\usepackage{mathrsfs}
\usepackage{bm}
\usepackage{slashed}
\usepackage{braket}
\numberwithin{equation}{section}
\usepackage{graphicx}
\usepackage{tikz}
\usepackage{tikz-cd}
\usepackage{subfig}
\usepackage{caption}
\usepackage[scaled]{couriers}
\usepackage{comment}

\usepackage{xcolor}
\usepackage{mdframed}

\usepackage[normalem]{ulem}
\usepackage{slashed}

\renewenvironment{figure}[1][]{
  \begin{originalfigure}[#1]
    \begin{mdframed}[linecolor=black!0,backgroundcolor=black!0]
}{
    \end{mdframed}
  \end{originalfigure}
}

\DeclareMathOperator{\Tr}{Tr}

\DeclareMathOperator{\Det}{Det}
\def\Re{\mathop{\mathrm{Re}}}
\def\Im{\mathop{\mathrm{Im}}}

\def\cA{{\cal A}}

\def\cC{{\cal C}}
\def\cD{{\cal D}}

\def\cJ{{\cal J}}

\def\cO{{\cal O}}

\def\cZ{{\cal Z}}

\def\bC{{\mathbb C}}

\def\bR{{\mathbb R}}

\def\bZ{{\mathbb Z}}

\def\sC{{\mathsf C}}
\def\sD{{\mathsf D}}

\def\sX{{\mathsf X}}
\def\sY{{\mathsf Y}}

\def\sc{{\mathsf c}}

\def\U{\mathrm{U}}
\def\SU{\mathrm{SU}}

\def\beq#1\eeq{\begin{align}#1\end{align}}

\def\d{{\rm d}}
\def\i{{\mathsf i}}

\def\CPN{\mathbb{CP}^{N-1}}
\def\energy{V}
\def\decay{\Gamma}
\def\vol{{\rm Vol\,}(T^2)}

\makeatletter
\@addtoreset{section}{app}
\makeatother

\begin{document}

\title{ Large $\theta$ angle in two-dimensional large $N$ $\mathbb{CP}^{N-1}$ model }

\preprint{TU-1255}

\author{Tsubasa Sugeno, Takahiro Yokokura, Kazuya Yonekura}
\affiliation{Department of Physics, Tohoku University, Sendai 980-8578, Japan }
\abstract{
In confining large $N$ theories with a $\theta$ angle such as four-dimensional $\mathrm{SU}(N)$ pure Yang-Mills theory,
there are multiple metastable vacua and it makes sense to consider the parameter region of ``large $\theta$ of order $N$'' despite the fact that $\theta$ is a $2\pi$-periodic parameter. We investigate this parameter region in the two-dimensional $\mathbb{CP}^{N-1}$ model by computing the partition function on $T^2$. When $\theta/N$ is of order $ \mathcal{O}(0.1)  $ or less, we get perfectly sensible results for the vacuum energies and decay rates of metastable vacua. However, when $\theta/N$ is of order $\mathcal{O}(1) $, we encounter a problem about saddle points that would give larger contributions to the partition function than the true vacuum. We discuss why it might not be straightforward to resolve this problem. 
}

\maketitle

\section{Introduction and Summary}

The theta angle $\theta$ in quantum field theories (QFTs) is an interesting parameter.
In particle physics and cosmology, the dependence of a theory on the theta angle is important for the potential energy of an axion or axion-like particles.
It is also interesting purely theoretically in QFTs because it appears only in nonperturbative phenomena (either instanton effects or strong coupling effects), and is related to topological properties of a theory.

Let us consider the case of four-dimensional (4d) $ \SU(N)$ gauge theories.
At weak coupling, we can calculate the $\theta$ dependence of the vacuum energy by the dilute instanton gas approximation.
As a result, the vacuum energy $\energy(\theta)$ as a function of  $\theta$ at the leading order of weak coupling expansion is given by $\energy(\theta) \propto (1-\cos\theta)$, where we have set the vacuum energy at $\theta=0$ to be zero.
However, the situation is different at strong coupling. 
To be concrete, let us focus on pure Yang-Mills theory. By large $N$ argument, it was found that the vacuum energy should be of the form $\energy(\theta) = N^2 \bar \energy(\theta/N)$, where $ \bar \energy$ is of order one in the large $N$ limit~\cite{Witten:1980sp,Witten:1998uka}.
\footnote{This difference in the behavior of the vacuum energy arises from the breakdown of the dilute instanton gas approximation at strong coupling, as pointed out by \citep{Witten:1978bc}.}
Moreover, even for finite $N$, an 't~Hooft anomaly was found that excludes the simple form $\energy(\theta) \propto (1-\cos\theta)$ with the assumption of a unique gapped vacuum at each $\theta$~\cite{Cordova:2019jnf,Cordova:2019uob} (see also \cite{Gaiotto:2017yup,Kikuchi:2017pcp}). 

A scenario that is consistent with both the large $N$ argument and the 't~Hooft anomaly is to have multiple metastable vacua.\footnote{In some theories such as ${\cal N}=1$ supersymmetric pure Yang-Mills theory, these vacua are exactly stable. } The $n$-th metastable vacuum has the vacuum energy $\energy_n(\theta) = N^2 \bar \energy( (\theta+2\pi n)/N)$. Including these metastable vacua, the vacuum energy has multi-branch structure in the sense that for a given $\theta$, there are multiple metastable vacua with energy $V_n(\theta)$. The $2\pi$-periodicity of $\theta$ is realized by a nontrivial monodromy $V_n(\theta+2\pi)=V_{n+1}(\theta)$.
Although the theory as a whole has the $2\pi$-periodicity, each branch (i.e. fixed $n$) is not $2\pi$-periodic.
This structure has applications to cosmology. For instance, natural inflation~\citep{Freese:1990rb} uses an axion-like inflaton whose potential is often assumed to be $\energy(\theta) \propto (1-\cos\theta)$. However, in strongly coupled theories, it is possible to realize more general potentials of the form $\energy_0(\theta) = N^2 \bar \energy( \theta/N)$ (see e.g.~\citep{Yonekura:2014oja,Nomura:2017ehb,Nomura:2017zqj}).

Recent numerical simulations for the case $N=2$ support the above prediction of the vacuum structure~\cite{Kitano:2020mfk,Kitano:2021jho,Yamada:2024vsk,Yamada:2024pjy,Hirasawa:2024fjt}.
In numerical simulations, however, it may not be easy (even if not impossible \cite{Kitano:2021jho}) to study the regime of metastable vacua that is far from the true vacuum. 
The true vacuum is given by $n$ such that $-\pi \leq \theta+2\pi n \leq \pi$. 
For cosmological applications, we may want to know the shape of the potential for large values of $\theta$ for a given $n$ (say $n=0$). For $\theta$ of order one, a general argument (see \cite{Witten:1998uka}) shows that $\energy_0(\theta) = N^2 \bar \energy( \theta/N) = \frac12 c \theta^2 + \cO(N^{-2})$ where $c = \bar \energy''(0)$ is positive~\cite{Vafa:1984xg}. For $\theta$ of order $N$ (i.e. $\theta/N$ is of order one), we need more detailed information about the function $\bar \energy$ for which there is no general prediction. This is the regime we are interested in this paper. 

For example, one qualitative question about the function $\bar\energy$ is whether $V_0(\theta)=N^2\bar\energy(\theta/N)$ has a $2\pi N$ (rather than $2\pi$) periodicity. The 't~Hooft anomaly mentioned above is an anomaly of the $2\pi$-periodicity of $\theta$. The anomaly vanishes for the $2\pi N$ shift of $\theta$. Therefore, it is consistent, but is not a general prediction, that $\energy_0(\theta)$ has a $2\pi N$-periodicity. See e.g. \cite{Yamazaki:2017ulc,Aitken:2018mbb,Tanizaki:2022ngt} for studies of some cases in which weak coupling computations are possible.

The two-dimensional (2d) $\CPN$ model is a good toy model which has several common properties with 4d pure Yang-Mills theory, such as asymptotic freedom, generation of mass gap, the existence of the $\theta$ parameter, and so on.
It also has a similar 't~Hooft anomaly about the $2\pi$ shift of $\theta$ as in the case of 4d Yang-Mills theory (see e.g. \cite{Nguyen:2022lie}).
Numerical simulations have also been performed in the 2d $\CPN$ model in many works~\citep{Campostrini:1992ar,Campostrini:1992it,Vicari:1992jy,Azcoiti:2003qe,Keith-Hynes:2008rvi,Laio:2015era,Rindlisbacher:2016cpj,Abe:2018loi,Bonanno:2018xtd} and some of them support the multi-branch structure of vacuum in this model~\citep{Azcoiti:2003qe,Keith-Hynes:2008rvi,Bonanno:2018xtd}.
Moreover, explicit computations are possible in the large $N$ limit~\cite{Witten:1978bc,DAdda:1978vbw,DAdda:1978dle}. 
We may get some insight into 4d Yang-Mills theory from the study of the $\CPN$ model.

The original motivation of the present paper is to study the vacuum energy $ \energy_n(\theta)$ of metastable vacua in the $\CPN$ model when $\theta$ is of order $N$. (However, we will encounter a puzzle which will be our main focus in this paper).
There are previous studies of the vacuum energy in the large $N$ $\CPN$ model.
The topological susceptibility (i.e. the constant $c$ in $ \energy_0(\theta) \simeq \frac12 c \theta^2$) has been determined in \cite{DAdda:1978vbw,Luscher:1978rn,Campostrini:1991kv}.
Higher orders of the expansion by $\theta$ are also determined in \cite{Lawrence:2012ua,Rossi:2016uce,Bonati:2016tvi,Bonanno:2018xtd,Berni:2019bch} by the method that we will also use in this paper. 
We calculate the partition function of the $\CPN$ model on $T^2$ which contains the information of the vacuum energy including metastable vacua.

The partition function $Z(\theta)$ on $T^2$ turns out to have a decomposition
\beq
Z(\theta) = \sum_{n \in \bZ} \cZ(\bar\theta_n), \qquad \bar\theta_n = \frac{\theta +2\pi n}{N}.
\eeq
The integer $n$ in this sum corresponds to the label of metastable vacua which appeared above.
As far as $|\bar\theta_n|$ is small (numerically of order $0.1$), we will get a sensible result $\cZ(\bar\theta_n) \sim \exp[ -  \energy_n(\theta) \vol ]$ where $ \energy_n(\theta) =N^2 \bar\energy(\bar\theta_n)$ and $\vol$ is the volume of $T^2$. More precisely, there is also an imaginary part so that $V_n(\theta)$ is replaced by $V_n(\theta) - \frac{\i}{2} \Gamma_n(\theta)$, where $ \Gamma_n(\theta)$ represents the vacuum decay rate of the $n$-th metastable vacuum. The result of $\bar\energy(\bar\theta_n)$ for small $|\bar\theta_n|$ agrees with previous results in the literature, and the vacuum decay rate will also have a clear interpretation as the Schwinger effect. 
For small $|\bar\theta_n|$, the absolute value $|\cZ(\bar\theta_n)|$ is a monotonically decreasing function of $|\bar\theta_n|$ and in particular $|\cZ(\bar\theta_n)| < |\cZ(0)|$ for $0 <|\bar\theta_n| < \cO(0.1)$.

However, when $|\bar\theta_n|$ is somewhat large (numerically of order $1$), we encounter a puzzle: {\it if we follow the standard large $N$ arguments such as the saddle point method, the result of computations is that the absolute value $|\cZ(\bar\theta_n)|$ seems to exceed $|\cZ(0)|$ for somewhat large $|\bar\theta_n| \sim \cO(1)$.} Recall that the true vacuum corresponds to $\bar\theta_n \sim 0$ (or more precisely $-\pi /N \leq \bar\theta_n  \leq \pi/N$). We expect that the true vacuum gives the most dominant contribution to the partition function in the large volume limit, and hence other contributions should not exceed the true vacuum contribution. Therefore, the above problem should be resolved in some way. We leave it a future work to find a resolution of the puzzle. In the present paper, we just present the puzzle and discuss why it might require something that is beyond the standard large $N$ analysis.

The rest of the paper is organized as follows. 
In Section~\ref{sec:preliminaries}, we will explain the basic setup.
Section~\ref{sec:effaction} gives a detailed review of the effective action in the large $N$ $\CPN$ model which is obtained after integrating out ``matter fields''.  
The $\CPN$ model is formulated by using an auxiliary $\U(1)$ gauge field as well as an auxiliary scalar field $D$, and the field strength $E$ of the $\U(1)$ gauge field is also a Lorentz scalar in 2d. The effective action is a function of $D$ and $E$. Although the effective action (including $E$) is known in the literature, we will try to be careful about the assumptions that are used to obtain the effective action.
In Section~\ref{sec:result}, we will show our main results. We evaluate the path integral by using the saddle points of the effective action. This is a standard procedure in large $N$ analysis.
First, for small $\bar\theta_n$ we will reproduce the known vacuum energy, and we will also compute the vacuum decay rate which is exponentially small for small $\bar\theta_n$.
We emphasize that the results for small $\bar\theta_n$ have perfectly sensible physical interpretations.
Next we will show numerical results for the saddle point actions for all $\bar\theta_n$.
By using those results, in Section~\ref{sec:puzzle} we describe the puzzle mentioned above, and explain why it might not be straightforward to resolve it.

\section{Preliminaries}\label{sec:preliminaries}

In this paper, we will perform the Euclidean path integral of the two-dimensional (2d) $\CPN$ sigma model in a large but finite  volume torus $T^2$. (The size of the volume will be discussed later.)  
In this section, we recall some basic facts about the $\CPN$ model and make some preparations for later sections.

\subsection{The basic setup}

The Euclidean action of the 2d $\CPN$ model with a $\theta$-term is given by
\begin{align}
  S_{\theta}[\phi,A,D] = \frac{N}{\bar g_0^2}\int \d ^2 x\ \left[ (D_\mu \phi)^\dagger (D^\mu \phi)+D(\phi^\dagger \phi -1) \right]
   - \i \frac{\theta}{2 \pi} \int \d ^2 x\  \frac{1}{2}\epsilon^{\mu\nu} F_{\mu\nu}, \label{eq:original-action}
\end{align}
where $\phi=(\phi^1,\cdots,\phi^N)$ is an $N$-component complex scalar field, $\bar g_0$ is a bare 't Hooft coupling constant, $D_\mu = \partial_\mu - \i A_\mu$ is a covariant derivative with an auxiliary $\U(1)$ gauge field $A_\mu$, $F_{\mu\nu}=\partial_\mu A_\nu-\partial_\nu A_\mu$ is the gauge field strength, and $D$ is an auxiliary scalar field. We will also use differential form notations such as $A = A_\mu \d x^\mu$ and $F = \d A$.
The Euclidean path integral is
\begin{align}
Z(\theta):=\int\mathcal{D}D\,\mathcal{D}A\,\mathcal{D}\phi^\dag\,\mathcal{D}\phi \, e^{-S_{\theta}[\phi,A,D]}.
\end{align}
Unless otherwise stated, the path integral is performed on a flat $T^2$.

The 2d $\CPN$ model has instanton sectors whose topological charge is given by 
\beq
m : =  \frac{1}{2 \pi} \int \d ^2 x\  \frac{1}{2}\epsilon^{\mu\nu} F_{\mu\nu} \in \bZ.
\eeq
We can divide the path integral into different topological sectors as
\begin{align}
Z(\theta)=\sum_{m \in\bZ}e^{\i m\theta}Z_m, \qquad
  Z_m:=\int_{m} \mathcal{D}D\,\mathcal{D}A\,\mathcal{D}\phi^\dag\,\mathcal{D}\phi\,e^{-S_{\theta=0}[\phi, A, D]}, \label{eq:instanton-sum}
\end{align}
where the subscript $m$ in the path integral for $Z_m$ means that it is restricted to the fixed topological sector specified by $m$.

In a large volume, it will be convenient to rewrite the sum over instanton numbers as follows~\citep{Aguado:2010ex}.
Suppose that $Z_m$, as a function of $m \in \bZ$, can be extended to a function $Z_x$ of a real variable $x \in \bR$. (This point will be discussed more explicitly later.)
Then, the Poisson summation formula $\sum_{m \in \bZ} \delta(x-m) = \sum_{n \in \bZ} e^{2\pi \i n x}$ gives

\begin{align}
Z(\theta)=\sum_{n\in\bZ}  \int_{-\infty}^\infty \d x \,e^{\i(\theta+2\pi n)x}Z_x. \label{eq:afterPoisson}
\end{align}
The integer $n$ is a kind of the electromagnetic dual of the 1-form gauge field $A_\mu$ in 2d.\footnote{The electromagnetic dual of a $p$-form gauge field in $d$-dimensions is a $(d-p-2)$-form. When $d=2$ and $p=1$, the dual is a ``$(-1)$-form'' which is just described by a constant integer $n \in \bZ$.}

The $n$ will be interpreted as labeling metastable vacua, at least when the vacuum decay rate is small.
We define the vacuum energy $\energy_{n}(\theta)$ and the vacuum decay rate $\decay_{n}(\theta)$ for the vacuum labeled by $n\in\bZ$ as 
\begin{align}
  \energy_{n}(\theta)-\i\frac{\decay_{n}(\theta)}{2}:=- \lim_{\vol \to \infty }\frac{1}{\vol}\log \left( \int_{-\infty}^\infty \d x \,e^{\i(\theta+2\pi n)x}Z_x \right).  \label{eq:energy-decay}
\end{align}
where $\vol$ is the volume of $T^2$.
We can recover the information of nonzero $n$ from that of $n=0$ because $\theta$ and $n$ appear in the combination $\theta +2\pi n$ and hence
\beq
 \energy_{n}(\theta)=\energy(\theta+2\pi n), \qquad \decay_{n}(\theta)=\decay(\theta+2\pi n), \label{eq:vacuumstr}
 \eeq
where $\energy(\theta)=\energy_{0}(\theta)$ and $\decay(\theta)=\decay_{0}(\theta)$.

\subsection{Large $N$ }

For large $N$ analyses, it is convenient to define
\beq
\bar \theta  = \frac{\theta}{N}.
\eeq
More generally, we will put a bar over a quantity that is normalized to be of order one in the large $N$ limit. 

We first perform the path integral over $\phi$ and $\phi^\dagger$. 
Then we obtain the effective action for $A_\mu $ and $D$ as
\begin{align}
  S_{\text{eff}}[A,D] =N \bar   S_{\text{eff}}[A,D] ,
  \label{eq:Seff1}
\end{align}
where 
\beq
\bar S_{\text{eff}}[A,D] =  \log \Det(-D_\mu D^\mu +  D) + \int \d ^2 x \left(   -\frac{1}{\bar g_0^2} D\,    - \i \frac{\bar \theta}{2 \pi}  \frac{1}{2}\epsilon^{\mu\nu} F_{\mu\nu} \right)     \label{eq:Seff1-1}
\eeq
This form of the effective action and the standard large $N$ argument suggest that the partition function is a function of $\bar\theta$ as
\beq
Z(\theta) \sim \exp\left( - N \bar S_{\text{eff}}(\bar\theta)  \right), \label{eq:nonperiodicZ}
\eeq
where $\bar S_{\text{eff}}(\bar\theta)$ is a function of $\bar\theta$ which is of order one in the sense of large $N$ counting. The contribution \eqref{eq:nonperiodicZ} is expected to be dominated by the vacuum. However, we know that the theory has the $2\pi$-periodicity $\theta \sim \theta+2\pi$. This periodicity would be inconsistent if \eqref{eq:nonperiodicZ} were the complete answer, since the function $\bar S_{\text{eff}}(\bar\theta)$ is of order one and the $2\pi$ shift of $\theta$ corresponds to the $2\pi/N$ shift of $\bar\theta=\theta/N$.  

The resolution is as follows. Suppose that there are many (metastable) vacua labeled by an integer $n \in \bZ$ such that the partition function is actually given by
\beq
Z(\theta) \sim \sum_{n \in \bZ} \exp\left( - N \bar S_{\text{eff}}(\bar\theta_n) \right) , 
\eeq
where
\beq
 \bar\theta_n := \frac{\theta+2\pi n}{N}. \label{eq:ntheta}
\eeq
Then it has the $2\pi$-periodicity.  This is a rather general structure of the $\theta$-dependence in large $N$ field theories. (For a nice summary in the case of Yang-Mills, see Section~1 of \cite{Witten:1998uka}.) For the $\CPN$ model, we have explicitly found this structure in \eqref{eq:afterPoisson} which is obtained after the Poisson summation.

\section{Effective action}\label{sec:effaction}
In this section, we study the effective action in the 2d $\CPN$ model. See \citep{Riva:1980wq,DAdda:1982lsk,Aguado:2010ex,Rossi:2016uce} for early work. We will try to make clear what assumptions we are using. 

\subsection{Gauge field configurations}\label{sec:gauge-configuration}
In a fixed topological sector labeled by $m \in \bZ$, we decompose the gauge field $A_\mu$ as
\beq
A_\mu = A_\mu^{(m)}+A_\mu^{\rm flat} + A'_\mu. \label{eq:gauge-decomposition}
\eeq
The meaning of each term on the right hand side is as follows. 
The $ A_\mu^{(m)} $ is a topologically nontrivial gauge field configuration such that the field strength is constant,
\beq
\frac12 \epsilon^{\mu\nu} F_{\mu\nu}^{(m)} = \frac12 \epsilon^{\mu\nu} \left( \partial_\mu A_\nu^{(m)} - \partial_\nu A_\mu^{(m)} \right)= \frac{2\pi m}{\vol}. \label{eq:gauge-topo}
\eeq
More explicitly, let us represent a torus $T^2$ (which we take to be rectangular for simplicity) by using a coordinate system $(x,y) \in \bR^2$ with equivalence relations
\beq
x \sim x+ L_1, \qquad y \sim y+L_2
\eeq
where $L_1$ and $L_2$ are the length of the two sides of $T^2$. We have $\vol =L_1L_2$.
Then $A^{(m)}$ may be taken as
\beq
A^{(m)}= A_\mu^{(m)} \d x^\mu =\frac{ 2\pi m}{L_1L_2}  x \d y, \label{eq:topological-gauge}
\eeq
whose field strength 2-form is $F^{(m)} = (2\pi m /L_1 L_2) \d x \wedge \d y$. We remark that there must be a nontrivial transition function $h(y)$ when we identify $x+L_1$ and $x$ because
\beq
-\i \left( A^{(m)}(x+L_1,y) -  A^{(m)}(x,y)  \right) = h(y)^{-1} \d h(y), \qquad h(y)=\exp\left(- \frac{2\pi m \i y}{L_2} \right).  \label{eq:transition}
\eeq
The $A_\mu^{\rm flat}$ and $A'_\mu$ are topologically trivial gauge fields describing fluctuations around the $A_\mu^{(m)} $. The $A_\mu^{\rm flat}$ is a flat connection of the form
\beq
A^{\rm flat} = 2\pi \left( \frac{\alpha  \d x}{L_1} + \frac{\beta \d y}{L_2} \right), \label{eq:flat-gauge}
\eeq
where $\alpha$ and $\beta$ are constants.
Because of large gauge transformations $-\i A \to -\i A + g^{-1} \d g$ by $\U(1)$-valued functions $g$, the parameters $\alpha$ and $\beta$ have gauge equivalence relations
\beq
\alpha \sim \alpha + 1, \qquad \beta \sim \beta + 1.
\eeq
The $A'_\mu$ contains other non-constant modes in the Fourier mode expansion of the (topologically trivial part of the) gauge field, or in other words the modes in $A'_\mu$ have nonzero momentum on $T^2$.

We remark that when $m \neq 0$, gauge field configurations with different values of $\alpha$ and $\beta$ are actually equivalent modulo translations on $T^2$ (up to some changes of $A'$ due to the translations). For instance, let us consider a translation $x \to x+L_1 a   $ by an arbitrary parameter $a$. Then
\beq
A^{(m)}(x+ L_1 a   , y) = A^{(m)}(x,y) + \frac{ 2\pi m a}{ L_2}   \d y.
\eeq
Thus, by choosing $a = - \beta /m$, we can cancel the part $2\pi \beta \d y/L_2$ in $A^{\rm flat}$. 
For a translation in the $y$-direction, we first perform gauge transformations by $g =\exp( 2\pi m \i xy/L_1L_2)$,
\beq
\tilde A^{(m)} =  A^{(m)} + \i g^{-1}  \d g=  - \frac{ 2\pi m}{L_1L_2} y \d x  .
\eeq
Then we can repeat the same argument as in the case of translations in the $x$-direction.\footnote{One might think that if we perform a translation of $A^{(m)}= \frac{ 2\pi m}{L_1L_2}  x \d y$ directly in the $y$-direction as $y \to y+L_2 b$, then we would get nothing because $\d (y+L_2b) = \d y$. However, it is not straightforward to do translations in the $y$-direction in this gauge, because we have the transition function $h(y)$ in \eqref{eq:transition} which is not invariant under translations in the $y$-direction. 
} 

\subsection{Large $N$ effective action}
Let us study the effective action \eqref{eq:Seff1-1}.
The first term in \eqref{eq:Seff1-1} can be rewritten as
\begin{align}
   \log \Det(-D_\mu D^\mu +  D) 
  =- \int_\epsilon^\infty\frac{\d t}{t} \Tr e^{-t(-D_\mu D^\mu +  D)} 
  \label{eq:logDetTr}
\end{align}
where we introduced a cutoff $\varepsilon$. 

The gauge field $A$ is decomposed as \eqref{eq:gauge-decomposition}.
Let us also decompose the auxiliary field $D$ as 
\beq
D = D_0 + D', \label{eq:D-decomposition}
\eeq
where $D_0$ is a constant on $T^2$ and $D'$ includes other non-constant modes in the Fourier mode expansion of $D$ on $T^2$ (i.e. they have nonzero momentum on $T^2$).
Then we define the effective action of $D_0$ and $m$ by
\beq
N \bar S_{\text{eff}}[m,D_0] = -\log\left( \int \mathcal{D} A^{\rm flat} \mathcal{D}A' \mathcal{D} D' e^{-N\bar S_{\text{eff}}[A,D]} \right) . \label{eq:partial-integration}
\eeq

For the time being, we focus on the case that $m \neq 0$. 
We {\it assume} that the integral over non-constant modes $A'$ and $D'$ is dominated by saddle points on which 
\beq
A'  , D' \sim \cO(e^{- M L_1})+\cO(e^{- M L_2}), \label{eq:A'D'vev}
\eeq
where $M = \sqrt{D_0}$ is the mass of the field $\phi$, and $L_1$ and $L_2$ are the lengths of the sides of $T^2$. Let us discuss the reasons for this assumption.

If we take the large volume limit of $T^2$, the spacetime is approximately the flat space $\bR^2$. Then, it may be reasonable to assume that the path integral is dominated by saddle points that are Lorentz and translation invariant, such as (metastable) vacua. If so, the values of $A'$ and $D'$ in those saddle points should vanish in the large volume limit because  $A'$ and $D'$ consist of modes with nonzero momentum. 

Now let us consider their behavior in the case of large but finite $L_1$ and $L_2$. If $ A^{(m)}$ were absent, translation invariance (i.e. momentum conservation) on $T^2$ would imply that the action $\bar S_{\text{eff}}[A,D]$ cannot have linear terms in $A'$ or $D'$ and hence is expanded as a function of $A'$ and $D'$ as
\beq
\bar S_{\text{eff}}[A,D] = \cO(1)   + \cO( A'^2) +\cO( D'^2) + \cO(A' D') + \cdots \label{eq:A'D'explansion1}
\eeq
where the ellipses denote quartic and higher order terms in $A'$ and $D'$. This would have $A'=0, D'=0$ as a saddle point. 

In reality, the translation invariance is violated by $ A^{(m)}$. However, the violation of translation invariance by $A^{(m)}$ only has the effect of changing $A^{\rm flat} $ as discussed at the end of Section~\ref{sec:gauge-configuration}. The flat connection $A^{\rm flat} $ has only global effects on $T^2$ because its field strength is zero, $\d A^{\rm flat}=0$. For it to be detected, a particle corresponding to the field $\phi$ must travel around one of the topologically nontrivial cycles of $T^2$. 
Then the quantum mechanical action of a $\phi$-particle going around the cycle of $T^2$ in the $x$-direction has the exponentiated action (see Appendix~\ref{app:particle})
\beq
\exp\left(- M \int \d s \right) \sim \exp(-ML_1) , \label{eq:particle-action}
\eeq
where $\d s$ is the line element of the worldline of the $\phi$-particle and $M$ is the mass.
Similarly, the exponentiated action of a $\phi$-particle going around the $y$-direction is given by $ \exp(-ML_2)$. This is the reason for the exponential suppression in \eqref{eq:A'D'vev}. See Appendix~\ref{app:exponential-terms} for more details.

At any rate, we assume the behavior \eqref{eq:A'D'vev} for the saddle point values of $A'$ and $D'$, and we {\it neglect} them by taking $L_1$ and $L_2$ to be sufficiently large.
Then let us consider the path integral on the right-hand side of \eqref{eq:partial-integration} around the saddle point $A' \sim 0, D' \sim 0$.
By the standard large $N$ argument, the $1/N$ can be regarded as a coupling constant in the perturbative expansion with respect to $A'$ and $D'$, and hence we expect
\beq
 \bar S_{\text{eff}}[m,D_0] = \bar S_{\text{eff} } [A=A_\mu^{(m)} ,D=D_0] + {\cal O}(1/N) +\cO(e^{- M L_1})+\cO( e^{- M L_2}) ,\label{eq:Seff-behavior}
\eeq
where we have used the fact that when $A' = 0$ and $ D' = 0$, the effect of $A^{\rm flat} $ can be eliminated by translation on $T^2$ as discussed at the end of Section~\ref{sec:gauge-configuration} and hence  $A^{\rm flat} $ does not affect the leading term of the effective action. (We must still perform the path integral over $A^{\rm flat} $, but it gives a constant overall factor of the partition function and is absorbed in the term $\cO(1/N)$ in \eqref{eq:Seff-behavior}.) We {\it assume} that the terms of order $ {\cal O}(1/N)$ as well as $\cO(e^{- c L_1}) +\cO( e^{-c L_2}) $ are negligible for the purposes of the following analysis.

Now we need to calculate $ \bar S_{\text{eff} } [A=A_\mu^{(m)},D=D_0] $. 
For this purpose, we use the standard result about Landau levels.
We denote
\beq
E_0 = - \frac{1}{2}\epsilon^{\mu\nu} F^{(m)}_{\mu\nu}  =  - \frac{2\pi m}{\vol}. \label{eq:E0-m}
\eeq
(The minus sign here is introduced just for later convenience.) 
We also denote
\beq
a_\pm = \frac{ \i( D_x \pm \i D_y) } {\sqrt{2|E_0|} },
\eeq
where $E_0 \neq 0$ since we are considering the case $m \neq 0$.
The $a_\pm$ are hermitian conjugates of each other and satisfy the commutation relation
\beq
[a_-, a_+]=-\frac{\i }{|E_0|} [D_x,D_y] = \frac{E_0}{|E_0|}.
\eeq
Depending on the sign of $E_0$, we can regard one of $a_\pm$ as an annihilation operator and the other as a creation operator. The numbers $N_\pm$ of modes annihilated by $a_\pm$ are determined by the Atiyah-Singer index theorem\footnote{The Dirac operator $\i \gamma^\mu D_\mu$ and the chirality operator $\bar \gamma$ in 2d are given by
\beq
\i \gamma^\mu D_\mu = \begin{pmatrix} 0 & \i( D_x - \i D_y) \\ \i( D_x + \i D_y) & 0\end{pmatrix}, \qquad \bar\gamma= \begin{pmatrix} 1 & 0 \\ 0 & -1\end{pmatrix}.
\eeq
} as
\beq
N_+ - N_- = \frac{1}{2\pi} \int \d x^2\, \frac{1}{2}\epsilon^{\mu\nu} F_{\mu\nu}  =m= -  \frac{1}{2\pi} \vol E_0.
\eeq
One of $N_\pm$ is zero because one of $a_\pm$ is a creation operator which does not annihilate anything. Then the other one of $N_\pm$ is given by $|m|$. 

Because $-D_\mu D^\mu = |E_0| (a_+ a_- + a_- a_+) $, its spectrum is the same as that of a harmonic oscillator in quantum mechanics. Actually, there are $|m|$ copies of harmonic oscillators because the lowest energy states have $|m|$-fold degeneracy. Thus we get
\begin{align}
  \Tr e^{-t(-D_\mu D^\mu +  D_0)}  = |m| \sum_{k=0}^\infty e^{- t (2k+1)|E_0|  - t D_0} 
  = \frac{1}{4\pi}\vol \ \frac{|E_0| e^{-D_0 t}}{\sinh |E_0| t} \, .
\end{align}
Then, the first term in (\ref{eq:Seff1-1}) becomes 
\begin{align}
     \log \Det(-D_\mu D_\mu +  D_0) 
    = -\dfrac{1}{4\pi} \vol \int_\varepsilon ^\infty \frac{\d t}{t}\  \frac{E_0 e^{-D_0 t}}{\sinh E_0 t} \, ,\label{eq:originalint}
\end{align}
where we have used the fact that $E_0/\sinh(E_0 t)$ is an even function of $E_0$.

Let us denote the effective action $\bar S_{\text{eff}}[m,D_0] $ defined in \eqref{eq:partial-integration} also as $\bar S_{\text{eff}}[E_0,D_0,\bar\theta] $, where $E_0$ and $m$ are related by \eqref{eq:E0-m}, and we have also made explicit the dependence on $\bar\theta$. From the above results, we get the effective action for $m \neq 0$ given by
\begin{align}
    \bar S_{\text{eff}}[E_0,D_0,\bar\theta] &\simeq \frac{1}{4\pi}  \vol \bar{\mathcal{L}}_{\text{eff}} (E_0,D_0,\bar\theta)  \,, \nonumber \\
    \bar{\mathcal{L}}_{\text{eff}}(E_0,D_0,\bar\theta) &:=  -   \int_\varepsilon^\infty \frac{\d t}{t} \frac{E_0 e^{-D_0 t}}{\sinh E_0 t} 
    + 2\i  \bar{\theta}E_0 - \frac{4\pi}{\bar g_0^2 }D_0  \, . \label{eq:effLag}
\end{align}
More explicit forms will be calculated in Section~\ref{effaspower}.

Let us recapitulate the {\it assumptions} used to arrive at the approximate expression \eqref{eq:effLag} for $ \bar S_{\text{eff}}[E_0,D_0,\bar\theta] $.
\begin{itemize}
\item The values of $A'$ and $D'$ in the dominant saddle points are exponentially suppressed as \eqref{eq:A'D'vev} and their effects are negligible.
\item Higher order terms in the $1/N$ expansion in \eqref{eq:Seff-behavior} are negligible. 
\end{itemize}

\subsection{The partition function}\label{sec:ptn}

Now the partition function is computed as follows. Recall the sum over instanton numbers in \eqref{eq:instanton-sum}. For $m \neq 0$, the $e^{\i m\theta}Z_m$ is given by $\hat Z[E_0,\bar\theta]$ define by
\beq
\hat Z[E_0,\bar\theta] = \int \d D_0 \, \exp\left(-N   \bar S_{\text{eff}}[E_0,D_0,\bar\theta] \right).
\eeq
We have computed the effective action only for $m\neq 0$. Thus, we rewrite \eqref{eq:instanton-sum} as
\beq
Z(\theta)= (Z_0-\hat Z[0]) + \sum_{m \in\bZ} \hat Z[E_0,\bar\theta].
\eeq
Now, one can see that the effective action \eqref{eq:effLag} (neglecting subleading terms) has a straightforward continuation from integer $m$ to real values, $m \to x \in \bR$. The $x$ and $E_0$ are related by
$
E_0 =  - {2\pi x /\vol}.
$
By using that continuation, we use the Poisson summation formula \eqref{eq:afterPoisson} to write the partition function as
\beq
Z(\theta) &= (Z_0-\hat Z[0]) + \sum_{n \in \bZ} \int \d x \, e^{2\pi \i n x} \hat Z[E_0,\bar\theta] \nonumber \\
  &=  (Z_0-\hat Z[0]) + \frac{\vol}{2\pi}  \sum_{n \in \bZ}   \int \d E_0 \d D_0  \, \exp\left(-N   \bar S_{\text{eff}}[E_0,D_0,\bar\theta_n] \right),
\eeq
where $\bar\theta_n = \bar\theta + 2\pi n/N$ was defined in \eqref{eq:ntheta}.
Thus, we have reduced the path integral to the integral over just two variables $E_0$ and $D_0$.

Let us briefly comment on $Z_0$ and $\hat Z[0] $. The $Z_0$ is (approximately) obtained by the following procedure. To get some intuition, we consider the case that $L_1=\beta=1/T$ is interpreted as an inverse temperature and $L_2  =L$ is a spatial volume which we take to be much larger than $\beta$ and the mass scale $M^{-1}=D_0^{-1/2}$ of $\phi$. In this limit, only the holonomy $\mu = \int_0^\beta A_x \d x$ in the Euclidean time direction $x$ is important. The path integral of $\phi$ gives
\beq
N \log \Det(-D_\mu D^\mu +  D)  = \beta L F(\mu),
\eeq
where $F(\mu)$ is the thermal free energy of $\phi$ in the presence of the (imaginary) chemical potential $\mu$. At the leading order, the $Z_0$ is given by
\beq
Z_0 \sim \int \d \mu \, e^{- \beta L F(\mu)} \sim e^{- \beta L F(0)}.
\eeq
The $\phi$ has mass $M $ and hence the thermal effects are exponentially suppressed by the Boltzmann factor $\exp(- \beta M)$ (which is the same as the exponentiated action \eqref{eq:particle-action} of the $\phi$-particle).

On the other hand, the $\hat Z[0]$ is computed by $  \bar S_{\text{eff}}[E_0,D_0,\bar\theta]$ in the limit $E_0 \to 0$, which one can see to be the zero temperature effective action of the $\CPN$ model for $D_0$. Thus, $Z_0$ and $\hat Z[0]$ are different when the temperature is of order $M$ or higher, but the difference goes to zero in the limit of large $L_1=\beta$ and $L_2=L$. 

Our concern in this paper is the $\bar\theta$ dependence, and for that purpose the term $ (Z_0-\hat Z[0]) $ is not important. We have also taken $L_1$ and $L_2$ to be large to justify the approximation used in \eqref{eq:effLag}. Therefore, at any rate, we neglect $ (Z_0-\hat Z[0]) $ and then our formula for the partition function is 
\beq
Z(\theta) \simeq  \frac{\vol}{2\pi}  \sum_{n \in \bZ}   \int \d E_0 \d D_0  \, \exp\left(- \frac{N \vol }{4\pi}     \bar{\mathcal{L}}_{\text{eff}}(E_0,D_0,\bar\theta_n) \right).\label{eq:pt-formula}
\eeq
The integration contours of $E_0$ and $D_0$ are as follows. It is clear that the $E_0$ is integrated over $\bR$. On the other hand, the $D$ was originally introduced as a Lagrange multiplier to impose the condition 
$|\phi|^2 = 1$. To produce the delta functional $\delta(|\phi|^2 - 1)$ by integrating over $D$ using the action \eqref{eq:original-action}, its integration contour should be pure imaginary, and hence $D_0$ is integrated over the imaginary axis $\sqrt{-1}\, \bR$. We may slightly shift it to $\epsilon+\sqrt{-1}\, \bR$ for $\epsilon >0$ to avoid the singular point $D_0=0$.

\subsection{Calculations of the effective Lagrangian}\label{effaspower}

The effective Lagrangian \eqref{eq:effLag} can be calculated more explicitly. To perform renormalization, we first rewrite
\beq
  \int_\varepsilon^{\infty} \frac{\d t}{t} \frac{E_0 e^{-t D_0}}{\sinh E_0 t} 
  =   \int_\varepsilon^{\infty} \frac{\d t}{t} \left(\frac{E_0 }{\sinh E_0 t} -  \frac{1}{t} \right)e^{-t D_0} +  \int_\varepsilon^{\infty} \d t \frac{e^{-t D_0}}{t^2} 
\eeq
The first term on the right-hand side is finite in the limit $\varepsilon \to 0$. On the other hand, the second term can be treated just as in the usual case of the $\CPN$ model without $E_0$ on a flat space.\footnote{Notice that
\beq
 \frac{1}{4\pi} \int_\varepsilon^{\infty} \d t \frac{e^{-t D_0}}{t^2} = \int \frac{\d^2 p}{(2\pi)^2}  \int_\varepsilon^{\infty} \frac{ \d t}{t} e^{-t(p^2+D_0)} = - \int \frac{\d^2 p}{(2\pi)^2} \log  (p^2 +D_0).
\eeq
This is just the flat space one-loop determinant. 
 } Because
\beq
\frac{\d^2}{\d D_0^2} \int_\varepsilon^{\infty} \d t \frac{e^{-t D_0}}{t^2} = \int_\varepsilon^{\infty} \d t e^{-t D_0}  = \frac{e^{- \varepsilon D_0} }{D_0} = \frac{1}{D_0}+\cO(\varepsilon), \label{eq:D0del}
\eeq
we have
\beq
 \int_\varepsilon^{\infty} \d t \frac{e^{-t D_0} }{t^2} = c_0 + c_1 D_0 + D_0 \log D_0 +\cO(\varepsilon),  \label{eq:D0del2}
\eeq
where $c_0 (= 1/\varepsilon) $ and $c_1 (=\log \varepsilon +\cdots)$ are constants which diverge in the limit $\varepsilon \to 0$. The $c_0$ is renormalized by the cosmological constant term and hence we neglect it. We renormalize $c_1$ as
\beq
c_1 + \frac{4\pi}{\bar g_0^2} : =-1- \log \Lambda^2,
\eeq
and take $\Lambda$ to be fixed in the limit $\varepsilon \to 0$. This $\Lambda$ is the dynamical scale of the $\CPN$ model.
By using this renormalization in \eqref{eq:effLag}, the renormalized effective Lagrangian is given by
\beq
    \bar{\mathcal{L}}_{\text{eff}}(E_0,D_0,\bar\theta) =  - D_0   \log (D_0/\Lambda^2)   + D_0
    + 2\i  \bar{\theta}E_0 -    \int_0^{\infty} \frac{\d t}{t} \left(\frac{E_0 }{\sinh E_0 t}  -  \frac{1}{t} \right)e^{-t D_0}    \, . \label{eq:effLag2}
\eeq
The term $ - D_0 \log (D_0/\Lambda^2) +D_0 $ is the usual one in the $\CPN$ model, and the rest of the terms gives the $E_0$-dependence. 

Next we calculate the last term of \eqref{eq:effLag2} more explicitly. We define a function $F(z)$ by
\beq
F(z) &:=  -\frac{1}{z} \int_0^{\infty}\left( \frac{1}{\sinh t} - \frac{1}{t} \right)\frac{e^{- z t}}{t}\d t  \nonumber \\
&= -\frac{2}{z} \log \Gamma\left( \frac{z+1}{2} \right) +  \log \left(\frac{z}{2} \right) -1 + \frac{1}{z}\log 2\pi , \label{eq:a-formula2}
\eeq    
where in the second equality we have used a formula \eqref{eq:a-formula} given in Appendix~\ref{app:formula}.
By setting 
\beq
z = \frac{D_0}{E_0},
\eeq
the final result for the effective Lagrangian when $E_0>0$ is given by
\beq
    \bar{\mathcal{L}}_{\text{eff}}(E_0,D_0,\bar\theta)  = - D_0 \log \left(\frac{ D_0}{\Lambda^2} \right)  + D_0 \left(1+ \frac{2 \i \bar \theta}{z} +F(z) \right) \qquad (E_0>0). \label{eq:full-formula1}
\eeq
More explicit form is
\beq
  \bar{\mathcal{L}}_{\text{eff}}(E_0,D_0,\bar\theta) = -2E_0 \log \Gamma\left( \frac{D_0}{2E_0}+\frac12 \right)   - D_0 \log \left(\frac{ 2E_0}{\Lambda^2} \right) +E_0 \log 2\pi +2\i \bar\theta E_0.\label{eq:full-formula2}
\eeq
The formula for negative real $E_0$ is given by replacing $E_0$ by $|E_0|$.
    
The above formulas in terms of the gamma function somewhat make obscure the behavior in the region of small $E_0$. It is possible to expand the effective Lagrangian around $E_0=0$ by using the Bernoulli numbers $B_n$ which are defined by 
\beq
\frac{x}{e^x-1} = \sum_{n=0}^\infty \frac{B_n}{n!} x^n.
\eeq
We have
\beq
\frac{x}{\sinh x}  = 2x \left(\frac{1}{e^x-1} -\frac{1}{e^{2x}-1} \right) = \sum_{n=0}^\infty  \frac{(2-2^{2n})B_{2n}}{(2n)!} x^{2n},
\eeq
where we have used the fact that the left-hand side is an even function of $x$. Then \eqref{eq:effLag2} gives
\begin{align}
  \bar{\mathcal{L}}_{\text{eff}}(E_0,D_0,\bar\theta) 
    = - D_0\left(\log\frac{D_0}{\Lambda^2} - 1 \right) +2\i \bar\theta E_0  -  \sum_{n=1}^\infty \frac{(2-2^{2n})}{2n(2n-1)} \, B_{2n} \, \frac{E_0^{2n}}{D_0^{2n-1}}\, . \label{eq:asymptotic}
\end{align}
The expansion with respect to $E_0$ is only an asymptotic expansion. In fact, the original integral \eqref{eq:effLag2} can be seen as a Borel resummation of this asymptotic series by rewriting it as
\beq
    \int_0^{\infty} \frac{\d t}{t} \left(\frac{E_0 }{\sinh E_0 t}  -  \frac{1}{t} \right)e^{-t D_0}  
    =   E_0 \int_0^{\infty} \frac{\d t}{t} \left(\frac{1 }{\sinh  t}  -  \frac{1}{t} \right)e^{-t \frac{D_0}{E_0}}   
\eeq
where we have assumed $E_0>0$. The function $ 1/\sinh t $ has poles at $\pi \i k$ for $k=\pm 1, \pm 2, \cdots$. Thus the radius of convergence of the expansion of $ 1/\sinh t $ is $\pi$, which implies that the expansion before the Borel resummation has zero radius of convergence. 

We can check that the results of this section are consistent with the previous studies~\citep{Riva:1980wq,DAdda:1982lsk,Aguado:2010ex,Rossi:2016uce}.
The effective action  \eqref{eq:effLag} in the present paper is equal to eq.~(3.4) in \citep{Riva:1980wq} and eq.~(3.5) in \citep{DAdda:1982lsk} up to the term derived from the fermionic superpartners of the bosonic fields.
Eq.~(32) in \citep{Rossi:2016uce} is equal to our result \eqref{eq:full-formula2} because $A$ and $B$ in \citep{Rossi:2016uce} are interpreted as $D_0$ and $E_0$ respectively in our case. 
In \citep{Aguado:2010ex}, the instanton number $k$ and the constant saddle point of the scalar field $m^2$ correspond to $E_0\mathrm{Vol}(T^2)/2\pi$ and $D_0$ respectively in our case.
Thus our result is equivalent to eq.~(41) in \citep{Aguado:2010ex} by replacing $2\pi k\rightarrow E_0\mathrm{Vol}(T^2)$ and $m^2\rightarrow D_0$.
Although the formulas are the same, we have tried to clarify the underlying assumptions for the derivations, including the estimate of finite volume effects in Appendix~\ref{app:exponential-terms}.

\section{The theta angle and the partition function}\label{sec:result}
Just for notational simplicity, from now on we denote $E_0$ and $D_0$ just as $E$ and $D$,
\beq
E:=E_0, \qquad D:=D_0.
\eeq
Our purpose now is to compute the partition function \eqref{eq:pt-formula}. For this purpose, we need to evaluate the integral 
\beq
 \cZ(\bar\theta)=\int \d E \d D  \, \exp\left(- \frac{N \vol }{4\pi}     \bar{\mathcal{L}}_{\text{eff}}(E,D,\bar\theta) \right), \label{eq:basic-integral}
\eeq
where we remind the reader that $\bar\theta=\theta/N$.
Then we replace $\bar\theta$ by $\bar\theta_n = \bar\theta + 2\pi n/N$ and sum over $n \in \bZ$ to get \eqref{eq:pt-formula}.

\subsection{The assumption about the saddle point method}\label{sec:saddle-method}
For the purpose of computing $\cZ(\bar\theta)$ defined by \eqref{eq:basic-integral}, we make a further {\it assumption} that it can be computed by the saddle point method. The saddle point method is a very common method in large $N$ field theories, but actually there is a subtlety. 

A precise framework for the saddle point method is Morse or Picard-Lefschetz theory (see Section~3 of \cite{Witten:2010cx} for a detailed review of this theory). An important ingredient in this method is the ability to deform the integration contour freely so that the integral is represented by a sum of integrals over some integration contours (called Lefschetz thimbles) which pass through saddle points. 

In more detail, let us extend the variables $E$ and $D$ to complex values, $E, D \in \bC$. The $  \bar{\mathcal{L}}_{\text{eff}}(E,D,\bar\theta) $ is now regarded as a function of these complex variables.
Suppose that $  \bar{\mathcal{L}}_{\text{eff}}(E,D,\bar\theta) $ has (complex) saddle points $p_a$ labeled by $a$, $p_a=(E_a,D_a)$. Let $\cJ_a$ be the appropriate integration contour (Lefschetz thimble) which has the following properties; (i)~$\cJ_a$ passes through $p_a$, (ii)~the real part $\Re   \bar{\mathcal{L}}_{\text{eff}}(E,D,\bar\theta) $ increases as we go away from $p_a$ along $\cJ_a$, and (iii)~the imaginary part $\Im   \bar{\mathcal{L}}_{\text{eff}}(E,D,\bar\theta) $ is constant along $\cJ_a$. Let $\cC$ be the original integration contour. Suppose that $\cC$ can be deformed to (or is in the same ``homology class'' as) a sum 
\beq
\cC \to \sum_a {\mathfrak n}_a \cJ_a,
\eeq
where ${\mathfrak n}_a $ are integers. Then we may want to replace the integral \eqref{eq:basic-integral} over $\cC$ as
\beq
&\int_\cC \d E \d D  \, \exp\left(- \frac{N \vol }{4\pi}     \bar{\mathcal{L}}_{\text{eff}}(E,D,\bar\theta) \right) \nonumber \\
 &\xrightarrow{~~{?}~~} \sum_a {\mathfrak n}_a \int_{ \cJ_a} \d E \d D  \, \exp\left(- \frac{N \vol }{4\pi}     \bar{\mathcal{L}}_{\text{eff}}(E,D,\bar\theta) \right). \label{eq:thimble-decomposition}
\eeq
The integral over each $\cJ_a$ is dominated by the saddle point value of the integrand. This is the saddle point method. 

This procedure is valid as far as the integrand is a holomorphic function of the integration variables $(E,D)$, because in that case we can use the Cauchy theorem to deform the integration contour without changing the value of the integral. However, as we have noticed in the paragraph containing \eqref{eq:asymptotic}, the $  \bar{\mathcal{L}}_{\text{eff}}(E,D,\bar\theta) $ is not analytic at $E=0$ because the radius of convergence of the expansion around this point is zero. The integration contour $\cC$ (discussed at the end of Section~\ref{sec:ptn}) passes through $E=0$. Therefore, it is not clear whether the decomposition such as \eqref{eq:thimble-decomposition} is possible.

In this paper, we neglect the above issue and simply {\it assume} the following: 
\begin{itemize}
\item The result for $\cZ(\bar\theta)$ is given by a sum that appears on the right-hand side of \eqref{eq:thimble-decomposition}, for some choice of integers ${\mathfrak n}_a$. (Any specific choice of ${\mathfrak n}_a$ is not assumed.)
\end{itemize}
 We will see that this assumption (with a certain choice of ${\mathfrak n}_a$) gives perfectly sensible results when $\bar \theta$ is small (numerically of order $0.1$). However, when $\bar \theta$ becomes larger (numerically of order $1$), then we will encounter a puzzling behavior. The puzzle will be that the values of $\Re   \bar{\mathcal{L}}_{\text{eff}}(E_a,D_a,\bar\theta) $ become negative and hence give unphysically huge contributions. This is not about a single saddle point but about many saddle points, and the problem might not be solved no matter how we choose ${\mathfrak n}_a$ (this point is subtle because there are infinitely many saddle points. We discuss more details later). 
 
 We leave the study of the resolution of the puzzle to future work. In the present work, we just give the results under the above assumptions.

\subsection{Small $\bar\theta$}\label{sec:small}
Let us start from the case of small $\bar\theta$ because this case has a very clear physical interpretation in terms of metastable vacua.
For small $\bar\theta$, we expect that small $ E $ gives the dominant saddle point, because when $\bar\theta=0$, the (true) vacuum is given by $E=0$.

For small $E$, the effective Lagrangian is expanded as
\begin{align}
  \mathcal{L}_{\text{eff}}(E,D,\bar\theta) 
    \simeq  2\i \bar\theta E -D\left(\log\frac{D}{\Lambda^2} - 1 \right)  + \frac{1}{6}\frac{E^{2}}{D} .
\end{align} 
Assuming that $\bar\theta$ is small, the saddle point is given by
\beq
D \simeq \Lambda^2+ 6\Lambda^2 \bar\theta^2, \qquad E \simeq -6\i \Lambda^2 \bar\theta, \qquad \mathcal{L}_{\text{eff}}|_{\rm saddle}   \simeq  \Lambda^2 + 6 \Lambda^2 \bar\theta^2. \label{eq:small-theta-saddle}
\eeq
Let us discuss the interpretation of each of them.

First, the leading term $ \Lambda^2$ in $D$ is the standard result for the $\CPN$ model at $\bar\theta=0$. This gives the mass term of $\phi$. The result $D=\Lambda^2$ for $\bar\theta=0$ is slightly modified by $6\Lambda^2 \bar\theta^2$ for a small but finite $\bar\theta$.

Next, let us consider $E \simeq -6\i \Lambda \bar\theta$. This is (almost) pure imaginary, and it has the following natural interpretation. The $E$ was defined by the gauge field strength $- \frac12 \epsilon^{\mu\nu} F_{\mu\nu}$ in Euclidean signature spacetime. We have to Wick-rotate it to obtain the corresponding Lorentz signature result. The Wick rotation is performed in the following way. The action contains the $\theta$ term
\beq
\frac{\i \theta}{2\pi} \int  E \d^2 x . \label{eq:fix-in-Wick}
\eeq
We require that this form is preserved under Wick rotation.\footnote{
We can write $\int  E \d^2 x$ in terms of differential form notation as $-\int F$. This does not depend on the metric. It is possible to regard Wick rotation as a change of the metric $g_{\mu\nu}$ (see \cite{Kontsevich:2021dmb} for a systematic study of this point of view), so Wick rotation should not affect terms that are independent of the metric. 
} 
Let $x_{\rm E}$ and $E_{\rm E}=E$ be the Euclidean signature coordinates and the field strength, and let $x_{\rm L}$ and $E_{\rm L}$ be the Lorentz signature coordinates and the field strength, respectively. Then the requirement is that $E_{\rm E}\, \d^2  x_{\rm E} = E_{\rm L}\, \d^2 x_{\rm L} $. We have the standard relation $\d^2 x_{\rm L} = -\i \d^2 x_{\rm E}$, and hence 
\beq
E_{\rm L} = \i E \simeq 6 \Lambda^2 \bar\theta. \label{eq:Loretnz-E}
\eeq
This is a real electric field. It is expected that such a constant electric field is generated in a 2d $\U(1)$ gauge theory with a $\theta$-term~\cite{Coleman:1976uz}.

Finally, the result $ \mathcal{L}_{\text{eff}}|_{\rm saddle} -\Lambda^2 \simeq  6 \Lambda^2 \bar\theta^2$ is interpreted in terms of the vacuum energy. From \eqref{eq:energy-decay}, the vacuum energy and the vacuum decay rate is given at the leading order of large $N$ by
\beq
\energy(\theta) = \frac{N}{4\pi} \bar \energy(\bar\theta), \qquad 
\decay(\theta) =  \frac{N}{4\pi} \bar \decay(\bar\theta)
\eeq
where
\beq
\bar \energy(\bar\theta) - \frac{\i}{2}\bar\decay(\bar\theta) =   \mathcal{L}_{\text{eff}} |_{\rm saddle} .
\eeq
Therefore, we get
\begin{align}
  \bar\energy(\bar\theta) -  \bar\energy(0) \simeq 6N\Lambda^2\bar\theta^2,
\end{align}
where we have subtracted the term $ \bar\energy(0)$ which is independent of $\bar\theta$. 
This is a natural result that the electric field $E_{\rm L}$ creates energy.

In the above approximation, we could not see the imaginary part of $  \mathcal{L}_{\text{eff}} |_{\rm saddle}$. We can determine the leading term of the small imaginary part as follows. We perform the above Wick rotation as $E = \exp[- \i (\pi/2 -\epsilon) ] E_L = (-\i + \epsilon) E_{\rm L}$, where $\epsilon>0$ is an infinitesimal number which is introduced so that the Wick rotation angle is slightly less than $\pi/2$ to avoid singularities in the following calculation.\footnote{This is analogous to the ``Feynman $\i \epsilon$'' in Feynman diagram computations. Our electric field $E$ is analogous to the time component of momenta in Feynman diagrams.} 
Then, the part of the effective action that depends on $E$ is
\beq
\mathcal{L}_{\text{eff}}  & \supset  -    \int_0^{\infty} \frac{\d t}{t} \left(\frac{E  }{\sinh E  t}  -  \frac{1}{t} \right)e^{-t D }  \nonumber \\
&=-    \int_0^{\infty} \frac{\d t}{t} \left(\frac{E_{\rm L} }{\sin (1 + \i \epsilon)E_{\rm L}  t}  -  \frac{1}{t} \right)e^{-t D } .
\eeq
When $E_{\rm L}$ is real (and $\epsilon$ is infinitesimal), there are many poles on the integration contour of $t$, located at
\beq
t =  \frac{\pi}{|E_{\rm L}|} k \qquad ( k=1,2,\cdots).
\eeq
Near these poles, we have
\beq
\frac{E_{\rm L} }{\sin (1 + \i \epsilon)E_{\rm L}  t} &\simeq  \frac{(-1)^k }{ (1 + \i \epsilon)  t -\pi k/|E_{\rm L}|}  \nonumber \\
&= {\rm P} \left( \frac{(-1)^k  }{   t - \pi k/|E_{\rm L}| } \right) -(-1)^k \pi \i \delta(   t - \pi k/|E_{\rm L}| ),
\eeq
where ${\rm P}$ means the principal value and we have used $(x + \i \epsilon)^{-1} = {\rm P}(x^{-1}) - \pi \i \delta(x)$. Therefore, for real $E_{\rm L}$, we get
\beq
\Im \mathcal{L}_{\text{eff}}   = \sum_{k =1}^\infty (-1)^k \frac{  |E_{\rm L}|}{k} \exp\left( - \frac{k\pi  D}{|E_{\rm L}|} \right).
\eeq
Therefore, when $|E_{\rm L}| \ll D$, the vacuum decay rate is given by
\beq
\bar\decay(\bar\theta) \simeq 2  |E_L| \exp\left( - \frac{ \pi D}{|E_{\rm L}|} \right) \simeq  12  \Lambda^2 |\bar\theta| \exp\left( - \frac{  \pi}{6|\bar\theta|} \right). \label{eq:schwinger}
\eeq
This is positive, as expected from the physical interpretation as the vacuum decay rate. In fact, if the Wick rotation were done in the wrong direction, then $\bar\decay(\bar\theta)$ would have been negative. The correct sign is a consistency check of our procedure for the Wick rotation discussed around \eqref{eq:fix-in-Wick}. 

The vacuum decay rate is exponentially small for small $\bar\theta$, and hence we can interpret the saddle point \eqref{eq:small-theta-saddle} as a sufficiently long-lived metastable vacuum.
The result \eqref{eq:schwinger} is interpreted as a consequence of the Schwinger effect; pairs of charged particles $\phi, \phi^\dagger$ are created in the presence of the electric field $E_{\rm L}$. The vacuum decay proceeds via such a Schwinger pair creation. In fact, in Appendix~\ref{app:schwinger} we give a different computation of the exponent $ -\pi D/|E_L|$ in \eqref{eq:schwinger} by using a particle picture of $\phi$. The computation there makes the interpretation as pair creations more explicit. 

\subsection{Analytic structure of the effective Lagrangian}\label{sec:analytic}
For general values of $\bar\theta$, we perform the computation by using the result \eqref{eq:full-formula1} or \eqref{eq:full-formula2}. If we regard $D$ and $z=D/E$ as independent variables, it is easy to solve the saddle point equation for $D$ by using \eqref{eq:full-formula1}. The result for the value of $D$ and the effective Lagrangian $  \bar{\mathcal{L}}_{\text{eff}}(z,\bar\theta) $ as a function of $z$ is
\beq
&D = D(z,\bar\theta):= \Lambda^2 \exp\left( \frac{2 \i \bar \theta}{z} +F(z)   \right), \nonumber \\
&  \bar{\mathcal{L}}_{\text{eff}}(z,\bar\theta) = D(z,\bar\theta), \label{eq:eff-z}
\eeq
where $F(z)$ is given by \eqref{eq:a-formula2},
\beq
F(z)= -\frac{2}{z} \log \Gamma\left( \frac{z+1}{2} \right) +  \log \left(\frac{z}{2} \right) -1 + \frac{1}{z}\log 2\pi . \label{eq:a-formula2-1}
\eeq

Our remaining task is to find saddle points of $ \bar{\mathcal{L}}_{\text{eff}}(z,\bar\theta) $. However, we must notice that this effective Lagrangian is not single valued for complex $z$. Moreover, the formula \eqref{eq:full-formula1} is derived under the condition that $E>0$. Therefore, it is necessary to be more careful about the analytic structure. 

Let us reexamine the case of small $\bar\theta$. The effective action was obtained for real $E$, but the actual saddle point for small $\bar\theta$ was approximately pure imaginary. In particular, to get the correct vacuum decay rate, we have implicitly used the following analytic continuation. We start from some real value of $E$ or $z=D/E$, say $z=A \in \bR$, and then perform Wick rotation as
\beq
z = A e^{\i \alpha} ,   \label{eq:howtocontinue}
\eeq
where $\alpha$ is increased from $0$ to $\pi/2$.
The sign of $A$ is taken to be the same as the sign of the saddle point value of $z_{\rm L} := D/E_{\rm L}$. 
For concreteness, we focus our attention to the case $\bar\theta>0$ which gives $E_{\rm L} \simeq 6 \Lambda^2 \bar\theta >0$. Then we take the $A$ in \eqref{eq:howtocontinue} to be positive. This means that we start the analytic continuation from positive $E$. Therefore we can use the formula \eqref{eq:full-formula1}. For $z>0$, the logarithm in \eqref{eq:a-formula2-1} is evaluated by the principal value. Then we analytically continue $F(z)$ along the path \eqref{eq:howtocontinue} for $A>0$. 

Now we want to increase $\bar\theta$ from small to large values. Let us distinguish several different types of saddle points as follows:
\begin{enumerate}
\item The saddle point that is a continuation of the saddle point found in Section~\ref{sec:small} from small to large values of $\bar\theta$. 
\item Saddle points that are not the above one, but can be still found by analytic continuation \eqref{eq:howtocontinue} in the range $0  < \alpha < \pi$ and $A>0$.
\item Saddle points that can be still found by analytic continuation \eqref{eq:howtocontinue} in the range $-\pi  < \alpha < 0$ and $A>0$.
\item Saddle points obtained by more complicated analytic continuation. 
\item Saddle points that are obtained by analytic continuation from negative real $z$. 
\end{enumerate}
Let us comment on each of them.

The first type is the most natural one because, for small $\bar\theta$, it has perfectly sensible behavior as a metastable vacuum. For large $\bar\theta$, however, we will see that the interpretation as a metastable vacuum is no longer valid. We will also see that the $\alpha$ in \eqref{eq:howtocontinue} must be continued to the range $\pi/2 < \alpha < \pi$. For small $\bar\theta$ it was $\alpha \simeq \pi/2$.

As we will see, there are also other saddle points. For the second type of saddle points, the analytic continuation is done in the same direction as the Wick rotation, $\alpha>0$. For the third type of saddle points, the analytic continuation is done in the opposite direction to the case of Wick rotation, $\alpha <0$. The saddle points of the third type (and the values of $\bar{\mathcal{L}}_{\text{eff}}(z,\bar\theta) $ on them) are obtained from the saddle points of the first or second types by complex conjugation $z \to z^*$. 

About the fourth type, a practical reason for distinguishing it from the first, second and third ones is as follows. The ``LogGamma function'' $ \log \Gamma(x)$ has a unique analytic continuation if we eliminate the negative real axis on the complex plane. The gamma function is basically a product of $(x+k)^{-1}$ for $k=0,1,2,\cdots$, or more precisely,
\beq
\Gamma(x)^{-1} = e^{\gamma x} x \prod_{k \geq 1} \left(1 + \frac{x}{k}\right)e^{-x / k}, \label{eq:productformula}
\eeq
where $\gamma$ is the Euler's constant.
For the function \eqref{eq:a-formula2-1}, if we go around some of the points of the form
\beq
z=-2k-1 \qquad (k=0,1,2\cdots), \label{eq:singularpoints}
\eeq
then the logarithm changes by $ -2\pi \ell \i$ for some $\ell \in \bZ$ and hence $F(z)$ changes as
\beq
F(z) \to F(z) - \frac{2}{z} \cdot 2\pi \ell \i.
\eeq
By looking at \eqref{eq:eff-z}, this shift of $F(z)$ has the effect of replacing $\bar\theta$ by $\bar\theta - 2\pi \ell$,
\beq
 \bar{\mathcal{L}}_{\text{eff}}(z,\bar\theta)  \to  \bar{\mathcal{L}}_{\text{eff}}(z,\bar\theta - 2\pi \ell). \label{eq:2piN?}
\eeq
Therefore, saddle points of the fourth type with the parameter $\bar\theta$ can be found from saddle points of the first, second or third type for the parameter $\bar\theta - 2\pi \ell$.\footnote{
It is not yet clear whether the ``$2\pi$-periodicity of $\bar\theta$'' or in other words the ``$2\pi N$-periodicity of $\theta$'' as in \eqref{eq:2piN?} has a physical interpretation. At least it is consistent with the 't~Hooft anomaly of the $2\pi$-periodicity of $\theta$~\cite{Cordova:2019jnf,Cordova:2019uob} (see also \cite{Gaiotto:2017yup,Kikuchi:2017pcp} for early work), since $2\pi N$ shift does not have any 't~Hooft anomaly.
}

The saddle points of the fifth type are obtained from others just by changing the signs of both $z$ and $\bar\theta$. Therefore we will not explicitly show them in the following sections.

\subsection{Numerical calculations}\label{sec:numerical}
\begin{figure}
\centering
\includegraphics[scale=0.45]{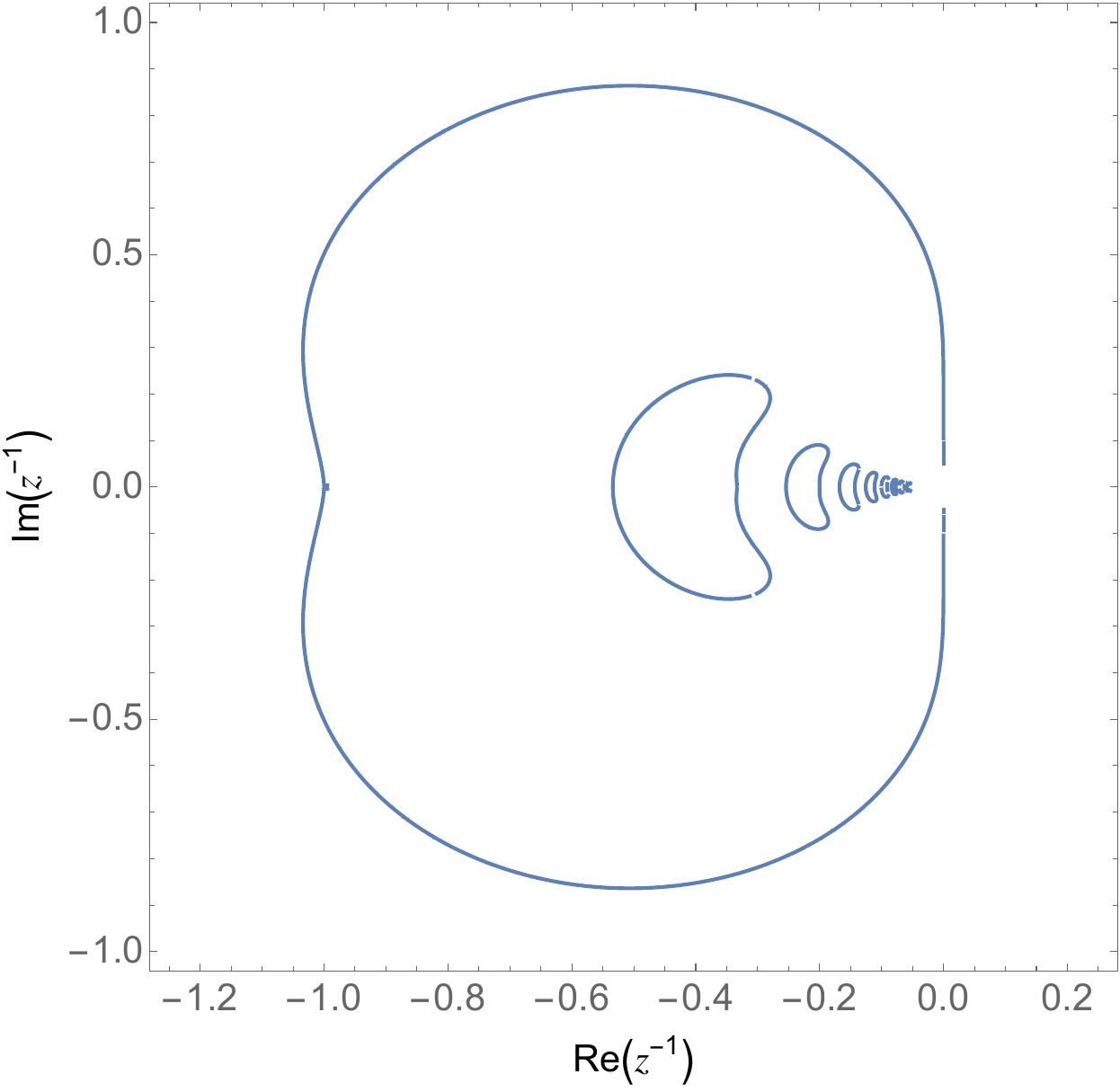}
\caption{The plot of saddle points on the complex $z^{-1}=E/D$ plane. The $\bar\theta$ is not fixed but is determined at each point by $\bar\theta=\frac12 \Im z^2 \frac{\d}{\d z} F(z)$. A single point corresponds to not a single value of $\bar\theta$ but multiple values which differ from each other by $2\pi \ell $ for $\ell \in \bZ$ (see the discussions around \eqref{eq:2piN?}). There are infinitely many ``rings'' as we approach $z^{-1} \to 0$ though it is not so clear in the figure due to numerical accuracy.  }
\label{fig:saddle}
\end{figure}

Now we determine the saddle points by numerical calculations. The saddle point equation for $z$ is given by
\beq
2 \i \bar\theta &=  z^2 \frac{\d}{\d z} F(z) \nonumber \\
&= 2\log \Gamma\left( \frac{z+1}{2} \right) - z \psi \left( \frac{z+1}{2} \right)  +z - \log 2\pi. \label{eq:z-saddle}
\eeq
where $\psi(x) = \frac{\d}{\d z} \log \Gamma(x)$ is the digamma function. This equation may be solved by imposing the condition $\Re z^2 \frac{\d}{\d z} F(z)  =0$. This gives curves on the complex plane. The value of $\bar\theta$ at each point on the curves is given by $\frac12 \Im z^2 \frac{\d}{\d z} F(z)  $.

Figure~\ref{fig:saddle}  shows the plot of the solutions of $\Re z^2 \frac{\d}{\d z} F(z)  =0$ on the complex plane of $1/z = E/D$.
The solutions form various ``rings''. Actually, there are infinitely many rings as we approach $z^{-1} \to 0$, though it is not so clear in the figure due to numerical accuracy. 

For small $\bar\theta$, we have a saddle point $z^{-1} \simeq - 6 \i \bar\theta$ found in Section~\ref{sec:small}. When $\bar\theta$ is increased, the point moves along the lower half of the largest ring in Figure~\ref{fig:saddle}. It reaches the point $z^{-1}=-1$ in the limit $\bar\theta \to \infty$. Notice that the point $z=-1$ is one of the singular points given by \eqref{eq:singularpoints}. In fact, each of the rings approaches one of the singular points $z = -2k-1$ in the limit $\bar\theta \to \infty$ (which will be discussed in Section~\ref{eq:approximate}). The upper half of Figure~\ref{fig:saddle} is obtained by complex conjugation $z \to z^*$ from the lower half.

\begin{figure}
  \centering
  \includegraphics[scale=0.42]{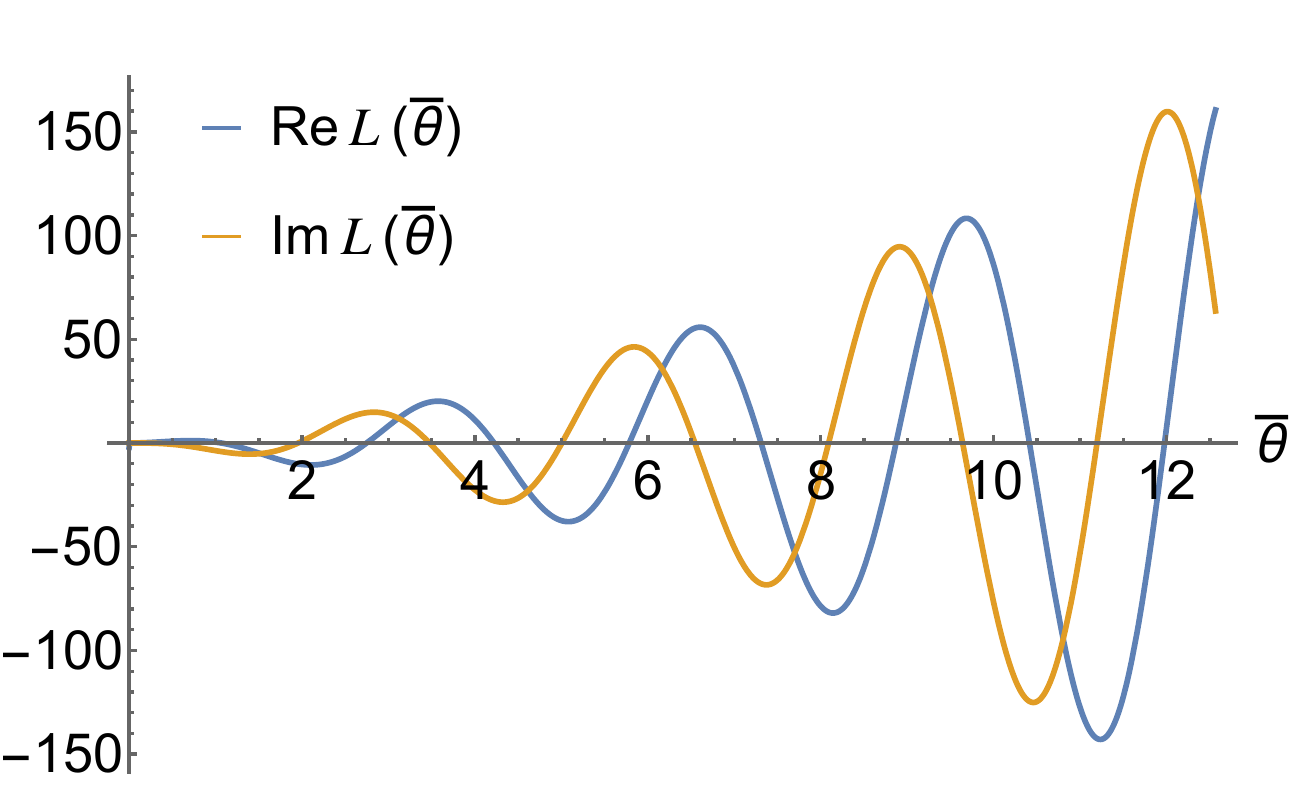}
  \caption{The real and imaginary parts of $L(\bar\theta)$ for the first ring ($k=0$). The blue curve is the real part and the orange curve is the imaginary part.} 
  \label{fig:vacuum}
\end{figure}

Let us next see the values of $ \bar{\mathcal{L}}_{\text{eff}}$ at the saddle points. The most natural saddle point is the one that is obtained continuously from the ones for small values of $\bar\theta$, that is, the lower half of the largest ring. 
Figure~\Ref{fig:vacuum} shows the result for 
\beq
L(\bar\theta)= \frac{\bar{\mathcal{L}}_{\text{eff}}|_{\rm saddle} -\Lambda^2}{\Lambda^2}. 
\eeq
Here, the constant $\Lambda^2$ is subtracted so that $L(0)=0$. A puzzling point is that the real part $\Re L(\bar\theta)$ becomes {\it negative} in some range of $\bar\theta$. We will discuss this point more in a later section.

\begin{figure}
  \centering
  \begin{minipage}{0.48\columnwidth}
    \centering
  \includegraphics[scale=0.3]{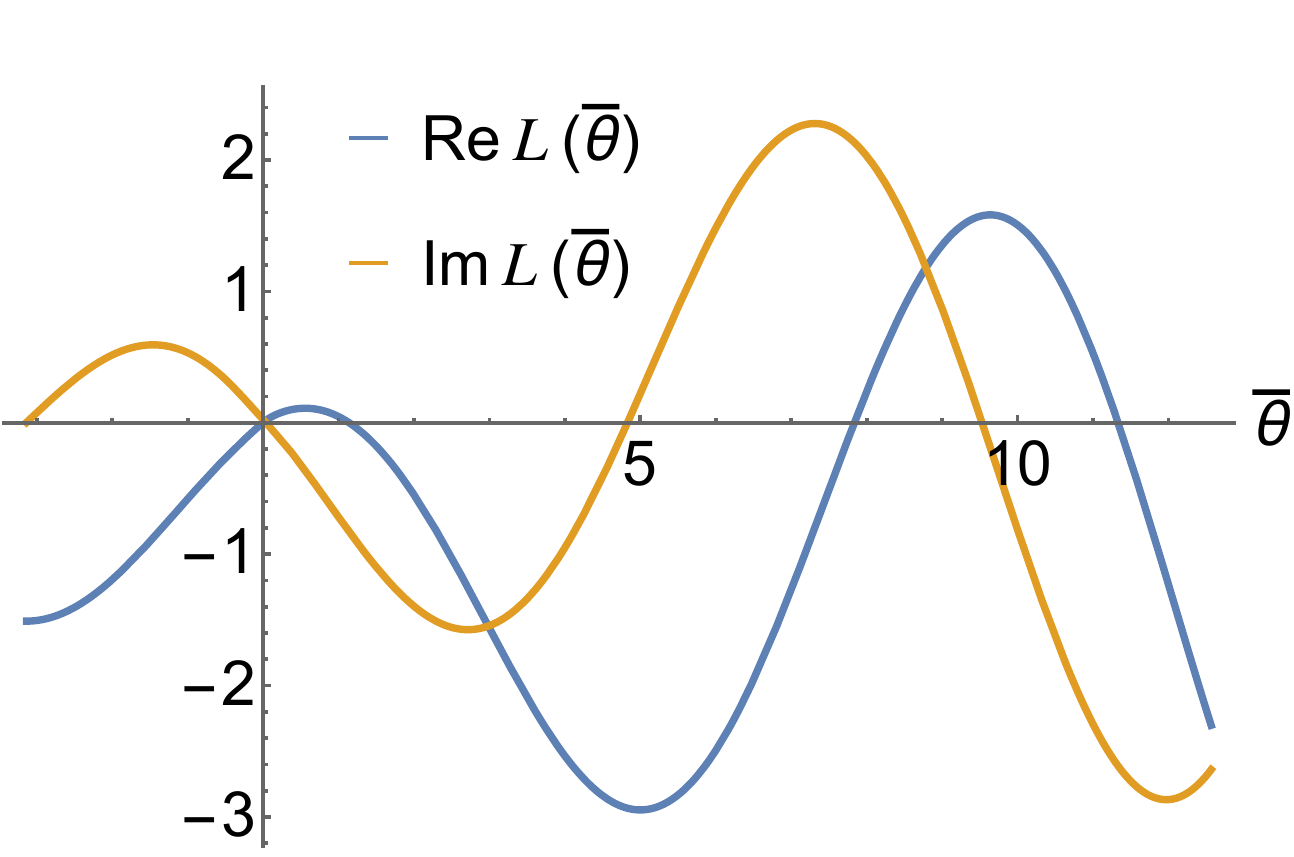}
\end{minipage}
\begin{minipage}{0.48\columnwidth}
  \centering
  \includegraphics[scale=0.3]{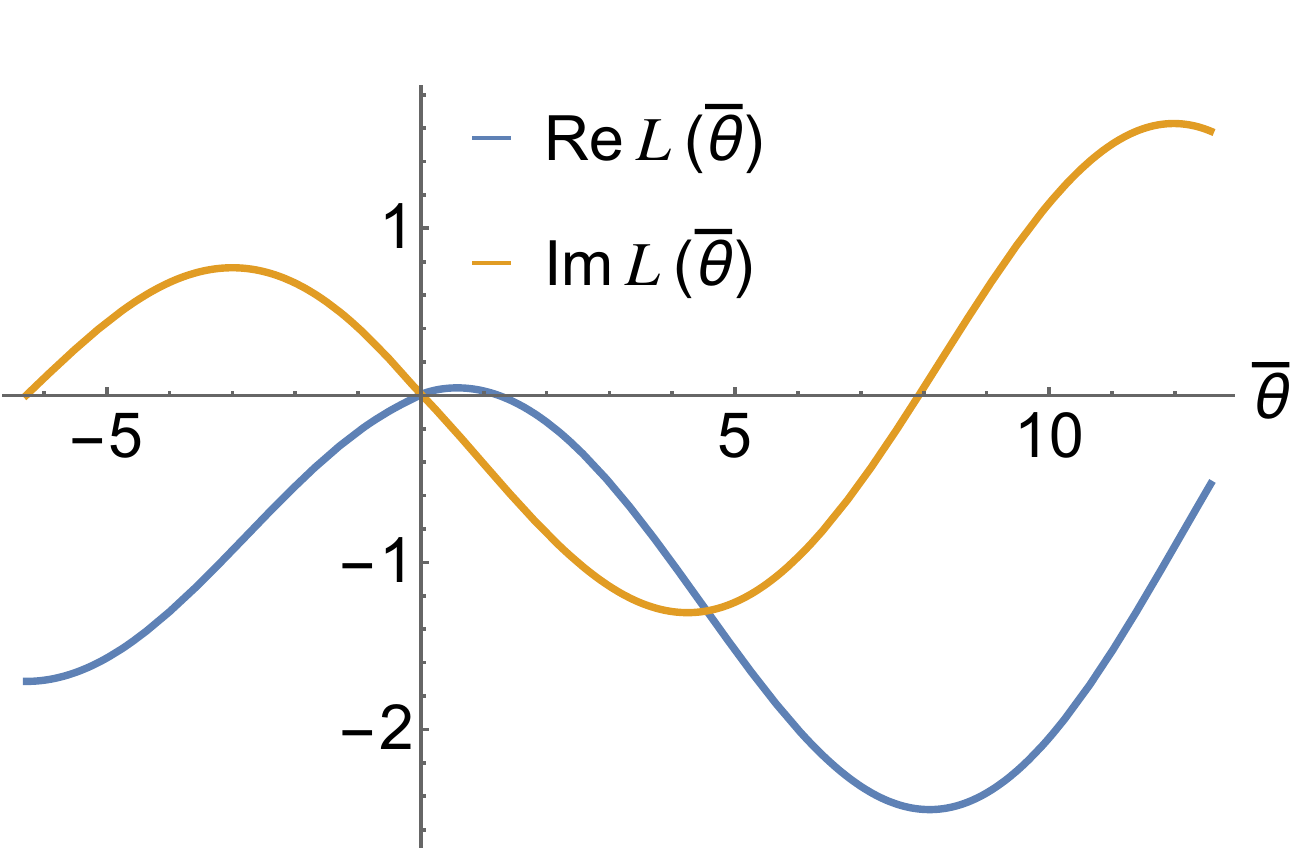}
  \end{minipage}
  \caption{The real and imaginary parts of $L(\bar\theta)$ for the second ring ($k=1$, Left) and the third ring ($k=2$, Right).} 
  \label{fig:vacuum2}
\end{figure}

\begin{figure}
  \centering
  \includegraphics[scale=0.34 ]{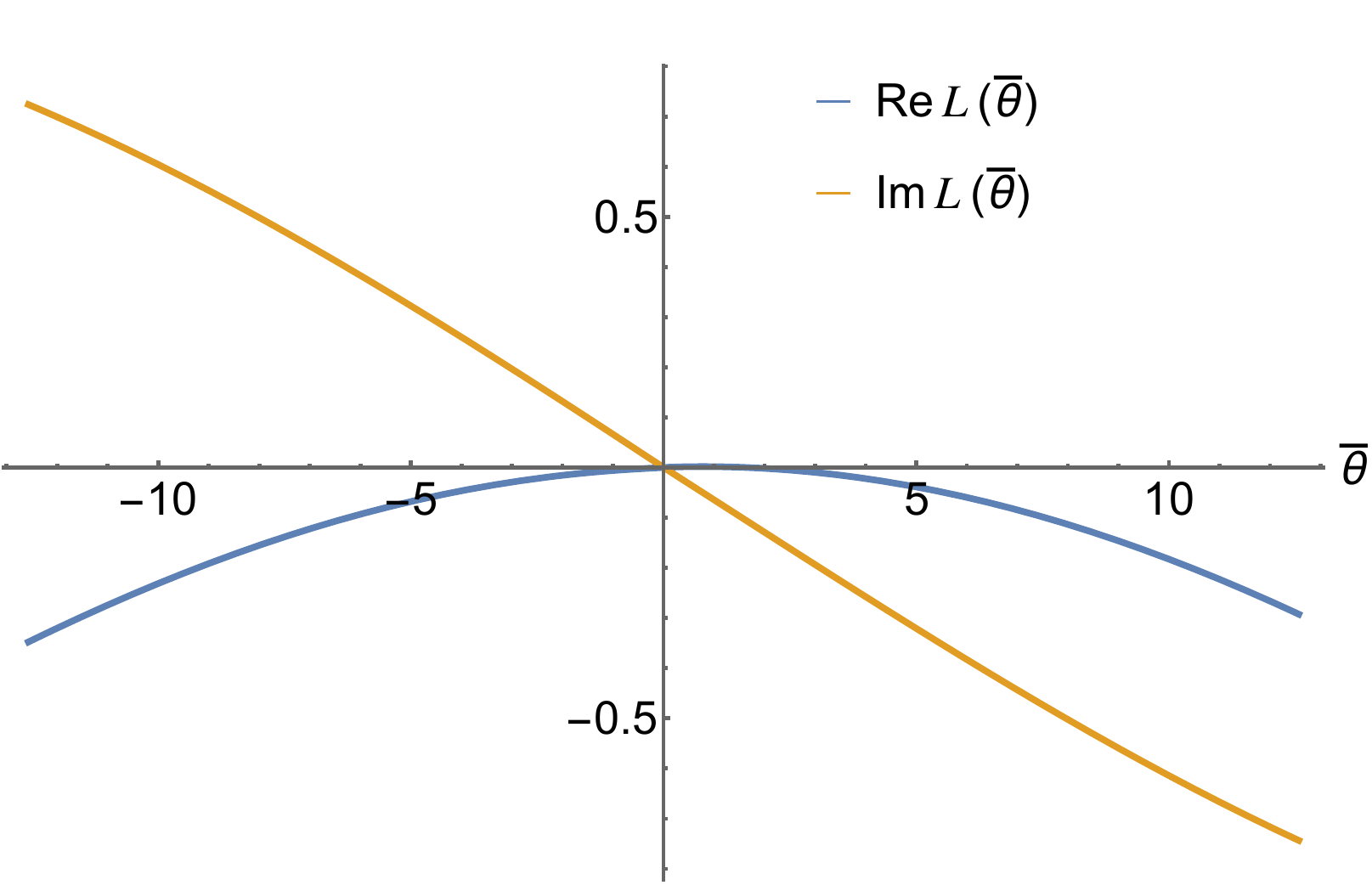}
  \caption{The real and imaginary parts of $L(\bar\theta)$ for the 15th ring ($k=14$). } 
  \label{fig:vacuum3}
\end{figure}

There are also other rings. Let us focus on the lower half of the rings. They have the property that the $(k+1)$-th ring behaves as $z^{-1} \to -(2k+1)^{-1}$ in the limit $\bar\theta \to \infty$ (see Section~\ref{eq:approximate}),
where $k=0$ is the one shown in Figure~\Ref{fig:vacuum}. The cases $k=1$ and $k=2$ are shown in Figure~\Ref{fig:vacuum2}. The case of a somewhat large $k$, e.g. $k=14$, is shown in Figure~\Ref{fig:vacuum3}. 

For these rings, we plot only the region 
\beq
\bar\theta > -\pi k \label{eq:thetarange}
\eeq
for the following reason. Let us consider analytic continuation \eqref{eq:howtocontinue} for $0 < \alpha <\pi$ and $-(2k+1) < A < -(2k-1)$. Then, in the limit $\alpha \to \pi$, one can see from \eqref{eq:z-saddle} and \eqref{eq:productformula} that $\bar\theta$ behaves as $\bar\theta \to - \pi k$. If we continuously change $\bar\theta$ to the region $\bar\theta < -\pi k$, then we go to the region of more complicated analytic continuation (i.e. the fourth type of saddle points in the terminology of Section~\ref{sec:analytic}). 
\footnote{There is a subtle defference in the case $k=0$. As we have dicussed in Section \ref{sec:analytic}, there are saddle points which are related to each other by the complex conjugation of $z$ and the sign flipping of $\bar\theta$. Only in the zeroth ring, these saddle points can be connected by continuously changing $\bar\theta$ to $-\bar\theta$ via $z^{-1}=0$. This means that if $\bar\theta$ is continuously changed to the region $\bar\theta <0$, the range of the analytic continuation shifts from $0<\alpha<\pi$ to $-\pi<\alpha<0$. In the terminology of Section~\ref{sec:analytic} about the type of saddle points, this is the transition from the second type to the third type. }
That region may be treated as discussed around \eqref{eq:2piN?}. 
It might be illuminating to write the equations for the saddle points as
\beq
&\i \bar\theta = (\log \Gamma)_{\rm pv} \left( \frac{z_{\rm saddle}+1}{2} \right) + 2\pi \ell \i - \frac{1}{2} z_{\rm saddle} \psi \left( \frac{z_{\rm saddle}+1}{2} \right)  + \frac{z_{\rm saddle}}{2} - \frac{1}{2} \log 2\pi , \nonumber \\
&\bar{\mathcal{L}}_{\text{eff}}|_{\rm saddle} =\frac{1}{2} \Lambda^2  z_{\rm saddle}     \exp\left(-  \psi \left( \frac{z_{\rm saddle}+1}{2} \right)  \right), \label{eq:saddlerep}
\eeq
where $z_{\rm saddle}$ is a saddle point value of $z$ which takes values on the rings of Figure~\ref{fig:saddle}, $ (\log \Gamma)_{\rm pv}$ is the ``LogGamma function'' defined by the unique analytic continuation in the region excluding the negative real axis, and $2\pi \ell \i $ represents the result of complicated analytic continuation beyond the negative real axis. The digamma function $\psi$ is single-valued, so the values of the action at saddle points are independent of $\ell$. The values of $\bar\theta$ at saddle points are affected by $\ell$, or in other words it is not uniquely determined by a point $z_{\rm saddle}$ on the rings. The explanation given above implies that the $ (\log \Gamma)_{\rm pv}$ has the property that $\bar\theta - 2\pi \ell > - \pi k$. 

The results for the upper half of the rings are obtained by complex conjugation and $\bar\theta \to -\bar\theta$. This region corresponds to $\alpha<0$ and hence they are opposite to the Wick rotation. 


From Figure~\Ref{fig:vacuum}, \Ref{fig:vacuum2} and \Ref{fig:vacuum3}, we find that ${\rm Re} L(\bar\theta)$ is negative for certain values of $\bar\theta$. For instance, as far as we can see from these figures, the ${\rm Re} L(\bar\theta)$  is negative for $\bar\theta \sim 1.5 $ (as well as $\bar\theta \sim 1.5 \pm 2\pi$ and $\bar\theta \sim - 1.5$) in all the figures. There are infinitely many saddle points and some of them may have positive ${\rm Re} L(\bar\theta)$. However, Figure~\Ref{fig:vacuum} and \Ref{fig:vacuum2} are the closest ones to the (continuation of the) saddle found in Section~\ref{sec:small}. It is unclear whether saddle points of the $(k+1)$-th ring for large $k$ or saddle points obtained by complicated analytic continuation \eqref{eq:2piN?} for large $|\ell|$ have any physical significance or not.

\subsection{Approximate solutions for large $\bar\theta$} \label{eq:approximate}

In the large $\bar\theta$ region, we have the oscillating behavior as can be seen in Figure~\ref{fig:vacuum} and \ref{fig:vacuum2}. It is possible to obtain approximate solutions for this region. 

Suppose that $z$ approaches one of the singular points \eqref{eq:singularpoints} and let us write
\beq
z = -(2k+1) +2 \i \xi,
\eeq
where $\xi$ is small. 
By analytic continuation \eqref{eq:howtocontinue} with $0 <\alpha <\pi$ (and $A>0$), we get
\beq
\log \Gamma\left( \frac{z+1}{2} \right) &= \log \Gamma\left( -k + \i \xi \right) \simeq  \log \left( \frac{ e^{-\i \pi k} }{k!} \cdot \frac{1}{\i \xi} \right) \nonumber \\
&=-\pi \i (k+1/2) - \log \xi - \log k!,
\eeq
where we have used $\Gamma(x+k) = (x+k-1)(x+k-2) \cdots x \Gamma(x)$ for $x=(z+1)/2$.

We will find that $\xi$ is of order $1/\bar\theta$. Thus we consider $\bar\theta \xi$ to be of order one, and then expand with respect to small $\xi$ to get,
\begin{multline}
 \frac{2 \i \bar \theta}{z} +F(z) \simeq - \frac{\i  \bar\theta }{k+1/2}\left( 1 +  \frac{\i \xi}{k+1/2} \right) 
-\frac{1}{k+1/2} \left(\pi \i (k+1/2) + \log \xi + \log k! \right)  \\
 + \log(k+1/2) + \i \pi -1 -\frac{1}{2k+1} \log 2\pi + \cO(\xi) .
\end{multline}
The equation of motion for $\xi$ gives
\beq
\xi \simeq \frac{k+1/2}{\bar\theta}.
\eeq
At this point we have
\beq
 \frac{2 \i \bar \theta}{z} +F(z) \simeq  \frac{1 }{k+1/2} \left( - \i  \bar\theta +\log \bar\theta \right) + C,
\eeq
where $C \in \bR$ is a real constant. The saddle point action $ \bar{\mathcal{L}}_{\text{eff}}|_{\rm saddle}$ is given by
\beq
 \bar{\mathcal{L}}_{\text{eff}}|_{\rm saddle} \simeq  e^C \Lambda^2 \exp\left(   \frac{- \i  \bar\theta +\log \bar\theta  }{k+1/2}   \right). \label{eq:approximateformula}
\eeq

\section{A problem about saddle points with huge contributions}\label{sec:puzzle}
In the previous section, we have found that the real part of the effective Lagrangian normalized as $ L(\bar\theta)=(\bar{\mathcal{L}}_{\text{eff}}|_{\rm saddle} -\Lambda^2)/\Lambda^2$ becomes negative for some values of $\bar\theta$. Now we discuss a puzzle about this behavior. 

\subsection{The puzzle}\label{sec:the-puzzle}
Modulo some overall factor that is irrelevant for our purposes, the partition function \eqref{eq:pt-formula}  is given by
\beq
Z(\theta) \propto \sum_{n \in \bZ} \cZ(\bar\theta_n), \qquad \bar\theta_n = \bar\theta + \frac{2\pi n}{N}, \label{eq:various-theta-sum}
\eeq
where $\cZ(\bar\theta)$ is defined by \eqref{eq:basic-integral}. Because we sum over $n \in \bZ$, the $\theta_n$ takes various values, some of which are large.

We have also assumed that the $\cZ(\bar\theta)$ can be computed by a sum of integrals over integration contours $\cJ_a$ (Lefschetz thimbles) as in \eqref{eq:thimble-decomposition}. Let $L_a(\bar\theta)$ be the value of $L(\bar\theta)$ at the saddle point labeled by $a$, and let $\cZ^{\rm loop}_a$ be the result of integration over $\cJ_a$ (without the saddle point action $L_a(\bar\theta)$). Then, modulo some irrelevant overall factor, we have
\beq
\cZ(\bar\theta) \propto \sum_a {\mathfrak n}_a \cZ^{\rm loop}_a \exp\left(- \frac{N \vol \Lambda^2}{4\pi} L_a(\bar\theta)  \right) ,  \label{eq:cZ}
\eeq
where the integers $ {\mathfrak n}_a$ are not yet known. We remark that $\cJ_a$ and $ {\mathfrak n}_a$ in general depend on $\bar\theta$, and may be written more precisely as $\cJ_a(\bar\theta)$ and $ {\mathfrak n}_a(\bar\theta)$. For notational simplicity, we omit writing the dependence on $\bar\theta$.

In our case, one representation of the label $a$ consists of two integers $k,\ell$ and a sign $o =\pm$ as follows.
In Section~\ref{sec:numerical}, we have discussed ``rings'' that appear in Figure~\ref{fig:saddle}. Then the integer $k$ specifies the $(k+1)$-th ring. For instance, Figure~\ref{fig:vacuum} is the result for $k=0$, and Figure~\ref{fig:vacuum2} is the result for $k=1$ (left figure) and $k=2$ (right figure). The sign $o =\pm$ specifies whether the analytic continuation \eqref{eq:howtocontinue} is done in the direction $0< \alpha < \pi$ which corresponds to the lower half of the rings, or $-\pi<\alpha<0$ which corresponds to the upper half of the rings. Also, we have the integer $\ell$ that appears in complicated analytic continuation as discussed around \eqref{eq:2piN?} and \eqref{eq:saddlerep}. Thus, we can write the label $a$ as $a=(k,\ell, o)$. We remark that
\beq
L_{(k,\ell,o)}(\bar\theta) = L_{(k,0, o)}(\bar\theta-2\pi \ell), \qquad L_{(k,0, -)}(\bar\theta) = L_{(k,0, +)}(-\bar\theta)^*. 
\label{eq:Lproperties}
\eeq
The $  L_{(k,0, +)}(\bar\theta)$ are the functions that are actually plotted in Figure~\ref{fig:vacuum}, \ref{fig:vacuum2} and \ref{fig:vacuum3} for $k=0,1,2$ and $14$.
By the reason explained in the paragraph containing \eqref{eq:thetarange}, we restrict $  L_{(k,0, +)}(\bar\theta)$ to $\bar\theta > -k \pi$. 
Therefore, the $\ell$ is restricted to $\bar\theta - 2\pi \ell > -\pi k$ or in other words $2\pi \ell <\bar\theta+\pi k$ in the representation $a=(k,\ell,o)$.

A different representation of the label $a$ is as follows. 
We can continue $L_{(k,0, +)}(\bar\theta)$ to the region $\bar\theta < -k \pi$ by ``going beyond the negative real axis'' in Figure~\ref{fig:saddle}. Then, as explained around \eqref{eq:thetarange}, the region $\bar\theta>-\pi k$ corresponds to the lower half of the rings in Figure~\ref{fig:saddle}, while the region $\bar\theta <-\pi k$ corresponds to the upper half of the rings.
Let $\tilde{L}_k$ be the function obtained by extending $L_{(k,0, +)}(\bar\theta)$ to $\bar\theta <-\pi k$.
By comparing analytic continuations from the lower half plane (i.e., $o=+$) and the upper half plane (i.e., $o=-$) of Figure~\ref{fig:saddle}, one can check that it is given by
\begin{align}
  \tilde{L}_{k}(\bar\theta) = 
     \begin{cases}
        L_{(k,0,+)}(\bar\theta) &\quad (\bar\theta>-\pi k), \\
        L_{(k,0,-)}(\bar\theta+2\pi k ) &\quad (\bar\theta<-\pi k),
     \end{cases}
\end{align}
The function $  \tilde{L}_{k}(\bar\theta)$ is now considered for all $\bar\theta$ without any restriction. 
For general $\ell$,  we simply replace the action by $  \tilde{L}_{k}(\bar\theta-2\pi \ell)$. Then the label $a$ is represented as $a=(k,\ell)$ without any restriction on $\ell$. 

The reason that there are two different representations of the label $a$ discussed above is that we can reach the upper half plane of Figure~\ref{fig:saddle} by either (i) the wrong Wick rotation (in the sense discussed in Section~\ref{sec:analytic}), or (ii) analytic continuation beyond the negative real axis. One can use whichever representation one wants to use. For concreteness, we use $a=(k,\ell,o)$.

We have not determined the integers ${\mathfrak n}_a$ in the partition function \eqref{eq:cZ}. First let us consider the case that for $(k,\ell, o)=(0,0,+)$ we have ${\mathfrak n}_{(0,0,+)}=1$. This is very natural for small $\bar\theta$, because it corresponds to the physically sensible metastable vacuum studied in Section~\ref{sec:small}. When we replace $\bar\theta \to \bar\theta_n$, we get the $n$-th metastable vacuum. However, as the value of $\bar\theta_n$ is increased, we encounter a problem. As observed in Section~\ref{sec:numerical}, the real part of ${L}_{(0,0, +)}(\bar\theta_n)$ becomes negative if $\bar\theta_n$ is moderately large (such as $\bar\theta_n \sim 1.5$). This means that, for a large $n$ such that $\Re {L}_{(0,0, +)}(\bar\theta_n)<0$, the contribution from such an $n$ would be huge in the sum \eqref{eq:various-theta-sum}. Physically we expect that the dominant contribution is the true vacuum, for which the value of $n$ is such that $\bar\theta_n \sim 0$ (or more precisely $-\pi \leq (\theta+2\pi n) \leq \pi$). Therefore, such a huge contribution is unlikely to be physical.

Even if ${\mathfrak n}_{(0,0,+)}=1$ for small $\bar\theta$, there is still a possibility that the value of ${\mathfrak n}_{(0,0,+)} = {\mathfrak n}_{(0,0,+)}(\bar\theta)$ jumps discontinuously at some value of $\bar\theta$ when this parameter is varied. Such jumps are known as Stokes phenomena (see Section~3 of \cite{Witten:2010cx} for a review). In Stokes phenomena, at least two saddle points are involved. If Stokes phenomena happen between two saddle points $(0,0,+)$ and $(k,\ell,o)$, the ${\mathfrak n}_{(0,0,+)}$ may change as 
\beq
{\mathfrak n}_{(0,0,+)} \to {\mathfrak n}_{(0,0,+)} + {\mathfrak m}\cdot {\mathfrak n}_{(k,\ell,o)}
\eeq
where ${\mathfrak m} \in \bZ$ is some integer.
Then there is a possibility to get ${\mathfrak n}_{(0,0,+)}=0$ for moderately large values of $\bar\theta$. However, for this mechanism to happen, we need $ {\mathfrak n}_{(k,\ell,o)} \neq 0$ for the saddle point $(k,\ell,o)$ that is involved in the Stokes phenomenon. It may be reasonable to speculate that Stokes phenomena happen between saddle points that are ``not too far from each other''. We have seen in Section~\ref{sec:numerical} that for not too large $k$ (such as $ k =0,1,2$) and $\ell=0, \pm 1$, the values of $\Re L(\bar\theta)$ for some $\bar\theta$ (such as $\bar\theta \sim 1.5$) are all negative. Thus the problem of a huge contribution happens if any of such ${\mathfrak n}_{(k,\ell,o)}$ is nonzero.

There are infinitely many saddle points and it is not possible to enumerate all of them. But we can study the cases that either $k$ or $|\ell|$ is large.
We have checked the behavior for large values of $k$. (For instance, Figure~\ref{fig:vacuum3} is for $k=14$.) 
For large and negative $\ell$, we may use the approximation in Section~\ref{eq:approximate} to see the behavior of ${L}_{(k,\ell,+)}(\bar\theta) = {L}_{(k,0, +)}(\bar\theta-2\pi \ell)$. For large and positive $\ell$, we use ${L}_{(k,\ell,-)}(\bar\theta) = {L}_{(k,\ell,+)}(2\pi \ell -\bar\theta)^*$.
In any case, as $k$ or $|\ell |$ (or both) becomes larger, the physical significance of the saddle point $(k,\ell,o)$ becomes more questionable.\footnote{
If both $k$ and $|\ell |$ are nonzero and one of them is not too small, it is possible to find a saddle point with $\Re {L}_{(k,0,+)}(\bar\theta -2\pi \ell)> 0$ by e.g.  using the formula \eqref{eq:approximateformula}. However, it is unclear whether such a saddle point is physically relevant.
}

\subsection{Discussions}

The resolution of the puzzle described above is not yet found. Here we would like to discuss some possibilities.

\paragraph{The validity of the saddle point method?} 
The precise treatment of the saddle point method (i.e. Morse or Picard-Lefschetz theory) assumes that the action can be extended to a holomorphic function of integration variables. As already mentioned in Section~\ref{sec:saddle-method}, our effective action is not holomorphic at $E=0$ and hence the validity of the saddle point method is questionable. We leave more detailed studies of this point for future work. If this is the resolution of the puzzle, our case gives an example in which the saddle point method is not applicable. Given the fact that the saddle point method is very commonly used in large $N$ field theories, we may gain new insights about large $N$ field theories from more studies of our example.

\paragraph{Nonzero momentum Fourier modes?} 
In the discussions around \eqref{eq:A'D'vev} as well as in Appendix~\ref{app:exponential-terms}, we have discussed that the $A'$ and $D'$ consisting of nonzero momentum Fourier modes on $T^2$ may have nonzero expectation values. The expectation values are exponentially small as $\exp(-\sqrt{D} L_1),~\exp(-\sqrt{D} L_2)$ where $L_1$ and $L_2$ are the lengths of the sides of $T^2$. These exponentials are small as far as the real part $\Re \sqrt{D}$ is positive. The equation \eqref{eq:eff-z} gives
\beq
\frac{  D }{\Lambda^2} =  L(\bar\theta)+1.
  \eeq
From the results given in Figure~\ref{fig:vacuum} (and \ref{fig:vacuum2} and \ref{fig:vacuum3}), we see that $\Re \sqrt{D}$ becomes negative when $\bar\theta$ is increased. 
In that region, our assumption of neglecting the expectation values of $A'$ and $D'$ may not be valid. However, we see from the figures that there are values of $\bar\theta$ such that we have both $\Re L(\bar\theta)<0$ and $\Re \sqrt{D} > 0$. For instance, in Figure~\ref{fig:vacuum}, $\bar\theta \sim 1.5$ is such a value. 

Even when $\exp(-\sqrt{D} L_1),~\exp(-\sqrt{D} L_2)$ are negligible, we have also assumed that the saddle points at which $A' \sim 0$ and $D' \sim 0$ are the most dominant ones. Logically there is a possibility that these saddle points do not contribute to the path integral, and instead, some saddle points that are not translationally invariant may contribute. However, recall that at least when $\bar\theta_n$ is small, we have a clear picture of metastable vacua for the saddle points with $A' \sim 0$ and $D' \sim 0$.

\paragraph{Stokes phenomena with complicated analytic continuation?} 
As mentioned in Section~\ref{sec:the-puzzle}, the situation is complicated due to the existence of infinitely many saddle points. We have already argued that simple Stokes phenomena between saddle points $(0,0,+)$ and $(k,\ell, o)$ for small $k $ and $|\ell|$ do not resolve the problem even if they really happen. However, we cannot exclude the logical possibility that some saddle points that are obtained by complicated analytic continuation (i.e., large $|\ell|$ and  $k$) are relevant. We leave more explicit study of Stokes phenomena for future work.

\paragraph{Cancellation in the sum over $n$?}
To obtain the final result for the partition function, we need to sum over all $n \in \bZ$ as in \eqref{eq:various-theta-sum}. Because of nonzero imaginary part $\Im L(\bar\theta)$, the phase of $\cZ(\bar\theta)$ is rapidly oscillating and there is a possibility that significant cancellation happens in the sum over $n$. 

For simplicity, we neglect several issues discussed above, such as Stokes phenomena.
When $n$ is changed to $n+1$, the exponent of the $\cZ(\bar\theta)$ in \eqref{eq:cZ} changes as
\beq
&\frac{N \vol \Lambda^2}{4\pi} \left( L_a(\bar\theta_{n+1}) -L_a(\bar\theta_{n})   \right)  \nonumber \\
&= \frac{\vol \Lambda^2}{2} \frac{\partial L_a}{\partial \bar\theta} (\bar\theta_n) + \frac{\pi \vol \Lambda^2}{2N} \frac{\partial^2 L_a}{\partial \bar\theta^2} (\bar\theta_n) + \cO(N^{-2}). \label{eq:random}
\eeq
Let us focus on the real part.
We have assumed that $\vol \Lambda^2 \gg 1$. If the first derivative $\Re ( \partial  L_a/ \partial \bar\theta)$ is nonzero and of order one, the difference of the real part of the action between $n$ and $n+1$ is very large. Then the term with lower value of the real part of the action dominates. It is difficult to have a cancellation between $n$ and $n+1$ because the absolute values of $\cZ(\bar\theta_n)$ and $\cZ(\bar\theta_{n+1})$ are very different (say $|\cZ(\bar\theta_n)| \gg |\cZ(\bar\theta_{n+1})|$ when $\Re ( \partial  L_a/ \partial \bar\theta)>0$). 
On the other hand, when $\bar\theta_n$ is such that $\Re ( \partial  L_a/ \partial\bar\theta) \sim 0$, the difference of the real part of the action is still large if $\vol \Lambda^2 \gg N$ and $\Re ( \partial^2  L_a/ \partial \bar\theta^2) \neq 0$. (A related point has been discussed in \cite{Aguado:2010ex} in the case that $\bar\theta_n \sim 0$.)

We did not make any assumption about the ratio $\vol \Lambda^2 / N$, so it is allowed to take a large $\vol$ such that $\vol \Lambda^2  \gg N$. 
In this limit, it seems difficult to have a cancellation in the sum over $n$. It would be interesting to study whether there is a loophole in this argument.

\acknowledgments

We would like to thank Yuya Tanizaki and especially Akikazu Hashimoto for helpful discussions.
We also thank Kai Murai and Yuma Narita for helpful advice on Mathematica.
 The work of KY is supported in part by JST FOREST Program (Grant Number JPMJFR2030, Japan), 
MEXT-JSPS Grant-in-Aid for Transformative Research Areas (A) ``Extreme Universe'' (No. 21H05188),
and JSPS KAKENHI (21K03546). 
The work of TS and TY is supported by Graduate Program on Physics for the Universe (GP-PU), Tohoku University.
TY is also supported by JST SPRING, Grant Number JPMJSP2114.

\appendix

\section{Particle picture}\label{app:particle}
The field $\phi$ of the model \eqref{eq:original-action} is a quantum field, but it is sometimes convenient to use particle quantum mechanics to study some effects of $\phi$. In this appendix we review the particle picture and discuss its applications. We treat $A_\mu$ and $D$ as background fields. 

Suppose we have a quantum mechanical particle in a spacetime (which is Euclidean $T^2$ or $\bR^2$ for our applications), coupled to a gauge field $A_\mu$. We parametrize the worldline by $\tau$, and the target space coordinate system is denoted by $x^\mu$. The line element $\d s$ and the pullback of the gauge field $A=A_\mu \d x^\mu$ are given by 
\beq
\d s^2 =\left(g_{\mu\nu} \frac{\d x^\mu}{\d \tau}\frac {\d x^\nu}{\d \tau} \right) \d\tau^2, \qquad A = A_\mu  \frac{\d x^\mu}{\d \tau} \d\tau,
\eeq
where $g_{\mu\nu}$ is the target space metric. Then the Euclidean action of the particle is given by
\beq
-S_{\rm particle} = -  \int  M\d s + \i \int A,\label{eq:p-action}
\eeq
where $M$ is the mass of the particle, which for the model \eqref{eq:original-action} is given by $M=\sqrt{D}$. 

\subsection{The relation between field and particle pictures}
Here we briefly review the relation between quantum field theory of $\phi$ and the particle theory \eqref{eq:p-action}. The relation will be given by
\beq
   -\log \Det(-D_\mu D^\mu +  M^2) = \int  \cD x^\mu \exp\left( - S_{\rm particle}  \right),  \label{eq:field-particle}
\eeq
where the right hand side is the path integral over particle trajectories $x^\mu(\tau)$ such that the worldline is topologically $S^1$.
 This is a well-known result, and for a more detailed review see e.g. \cite{Witten:2015mec}. Readers who can just accept the result can skip this subsection.  

We have the formula \eqref{eq:logDetTr}, which we repeat here with a slight change of variable $t$ by a factor of 2 and also a replacement $D \to M^2$:
\beq
   -\log \Det(-D_\mu D^\mu +  M^2) 
  = \int \frac{\d t}{t} \Tr e^{-\frac12 t(-D_\mu D^\mu +  M^2)} .   \label{eq:logDetTr-1}
\eeq
Now, we regard the derivative operator $H = \frac12 t(-D_\mu D_\mu +  M^2)$ as a Hamiltonian of a quantum mechanical particle. Namely, if we start from an action (which is written in Euclidean signature time $\tau$ on the worldline)
\beq
-\widehat{S}_{\rm particle} = - \int \d \tau \left( \frac{1}{2t} g_{\mu\nu} \frac{\d x^\mu}{\d \tau}\frac {\d x^\nu}{\d \tau}  + \frac12 t M^2\right)+ \i \int A,
\eeq
then its quantization gives the above Hamiltonian $H$.
The $\Tr e^{-H}$ is computed by the path integral using this action $\widehat{S}_{\rm particle}$ on a circle $S^1$. We take $\tau$ to have values in $[0,1]$ with the two ends $\tau=$0 and $\tau=1$ identified to make $S^1$, since we are computing $\Tr e^{-\beta H}$ with $\beta=1$. (For more general $\beta$, the $\tau$ would take values in $[0,\beta]$.)

We can also interpret the integration variable $t$ in \eqref{eq:logDetTr-1} geometrically. Consider a worldline metric $e^2 \d \tau^2$ where $e^2$ is the metric tensor on the (Euclidean signature) worldline. The only diffeomorphism-invariant information of $e$ on $S^1$ is the length of $S^1$ which we write as
\beq
t = \int_0^1 e \d \tau.
\eeq
We interpret the parameter $t$ in \eqref{eq:logDetTr-1} as this length.
Then the $\widehat{S}_{\rm particle}$ is the gauge-fixed version of the action
\beq
-\widehat{S}_{\rm particle} = - \int \d \tau\, e \left( \frac{1}{2} e^{-2} g_{\mu\nu} \frac{\d x^\mu}{\d \tau}\frac {\d x^\nu}{\d \tau}  + \frac12  M^2\right)+ \i \int A,
\eeq
where the gauge fixing is given by $e=t$.
The integral $\int \frac{\d t}{t}$ is interpreted as the path integral measure of $e$.\footnote{We do not try to give a full explanation of the fact that the integral measure is $\frac{\d t}{t}$ rather than just $\d t$. It is related to the fact that, after gauge fixing $e=t$, there are remaining diffeomorphisms which are translations on $S^1$ and hence we need to divide by its gauge volume. For a fuller explanation, readers may consult textbooks on string theory.} 

Combining the above discussions, we get the formula
\beq
   -\log \Det(-D_\mu D_\mu +  M^2) = \int \cD e \cD x^\mu \exp\left( -\widehat{S}_{\rm particle}  \right),
\eeq
where the right hand side is a path integral over $e$ and $x^\mu$ on the worldline $S^1$. The $e$ does not have derivative terms and hence we may just integrate it out. Its value on the saddle point is given by
\beq
e = \frac{1}{M} \sqrt{g_{\mu\nu} \frac{\d x^\mu}{\d \tau}\frac {\d x^\nu}{\d \tau} }.
\eeq
By substituting it into $\widehat{S}_{\rm particle}$, the action now becomes \eqref{eq:p-action} and we finally arrive at the formula \eqref{eq:field-particle}.

\subsection{Some terms in the effective action}\label{app:exponential-terms}
In this subsection, we give a rough estimate of some terms in the effective action \eqref{eq:Seff1-1} of $A$ and $D$,
\beq
\bar S_{\text{eff}}[A,D] =  \log \Det(-D_\mu D^\mu +  D) + \int \d ^2 x \left(   -\frac{1}{g_0^2} D\,    - \i \frac{\bar \theta}{2 \pi}  \frac{1}{2}\epsilon^{\mu\nu} F_{\mu\nu} \right)     .\label{eq:Seff1-2}
\eeq 
Recall the decompositions \eqref{eq:gauge-decomposition} and \eqref{eq:D-decomposition},
\beq
A_\mu = A_\mu^{(m)}+A_\mu^{\rm flat} + A'_\mu, \qquad D = D_0 + D'.
\label{eq:decomposition}
\eeq
The $A'$ and $D'$ consist of Fourier modes with nonzero momentum on $T^2$. We want to study terms in the effective action that are linear in $A'$, $D'$. As discussed around \eqref{eq:A'D'explansion1}, such linear terms would be forbidden by momentum conservation if $ A_\mu^{(m)}$ were absent. The $ A_\mu^{(m)}$ is not translationally invariant and hence it can break momentum conservation. However, translations of $ A_\mu^{(m)}$ only have the effect of shifting $A_\mu^{\rm flat}$ as discussed in Section~\ref{sec:gauge-configuration}. Therefore, the violation of momentum conservation must be sensitive to $A_\mu^{\rm flat}$. The flat connection $A_\mu^{\rm flat}$ is a global effect on $T^2$ since its field strength is zero, $\d A^{\rm flat}=0$.

Among topologically nontrivial trajectories, the ones that minimize the real part of the action $\Re S_{\rm particle}$ are given by straight lines in the $x$ or $y$ directions on $T^2$. Let $\sX_y$ and $\sY_x$ be trajectories on $T^2$ given by
\beq
\sX_y =\{ (\tau, y)~|~ 0 \leq \tau \leq L_1\}, \qquad \sY_x =\{ (x, \tau)~|~ 0 \leq \tau \leq L_2\}.
\eeq
Thus, the $\sX_y$ is a cycle in the $x$-direction with a fixed $y$, and the $\sY_x$ is a cycle in the $y$-direction with a fixed $x$. The holonomies of the gauge field along them are given by
\beq
\exp\left(\i \int_{\sX_y} A \right) &= \exp \left[ 2\pi \i \,(-m  y/L_2 +  \alpha ) \right] \left(1+ \i \int \d x A'_x + \cO(A'^2)\right), \nonumber \\
\exp\left(\i \int_{\sY_x} A \right) &= \exp\left[ 2\pi \i\,  (m  x/L_1 +  \beta) \right] \left(1+ \i \int \d y A'_y + \cO(A'^2)\right), 
\eeq
where we have used the gauge field configuration \eqref{eq:topological-gauge} for the case $ \sY_x$, the transition function \eqref{eq:transition} for the case $\sX_y$,\footnote{
More precisely, the meaning of the transition function \eqref{eq:transition} is as follows. The $T^2$ has the universal cover $\bR^2$ whose coordinate system is $(x,y) \in \bR^2$.
The gauge bundle can be trivialized on the $\bR^2$, and such a trivialization is implicitly assumed when we write the explicit gauge field \eqref{eq:topological-gauge}. 
We parametrize the bundle on $\bR^2$ by coordinates $(x,y,v) \in \bR^2 \times \bR$, where $v$ is the coordinate of the bundle (in the defining representation of $\U(1)$). Now, let us go back to $T^2$. 
We identify two elements $(x,y,v_{(x,y)} ) \in \bR^2 \times \bR$ and $(x+L_1, y,v_{(x+L_1,y)} ) \in \bR^2 \times \bR$ by using the transition function as $(x, y, v_{(x,y)}) \sim (x+L_1, y, h(y)^{-1} v_{(x,y)})$ or in other words $v_{(x,y)} = h(y) v_{(x+L_1,y)} $. (We also identify $(x, y, v_{(x,y)}) \sim (x, y+L_2,  v_{(x,y)})$ or in other words $v_{(x,y)} = v_{(x,y+L_2)}$.)
One can check that the identification is consistent with the covariant derivative $D_\mu{(x,y)}=\partial_\mu -\i A_\mu(x,y)$ and the relation \eqref{eq:transition}, i.e., $D_\mu{(x,y)} v_{(x,y)} = h(y) D_\mu{(x+L_1,y)}  v_{(x+L_1,y)}$.

With the above understanding, the holonomy of $A^{(m)}$ along $\sX_y$ is computed as follows. We start from some element $(0,y,v) \in \bR^2 \times \bR$. Because the $x$-component $A_x^{(m)}$ is zero, the parallel transport of the element along $\sX_y$ simply gives $(L_1, y, v)$. This is identified with $(0,y, h(y)v)$. Thus the holonomy of $A^{(m)}$ along $\sX_y$ is given by $h(y) =  \exp [ 2\pi \i \,(-m  y/L_2  ) ]$.
} and the flat connection \eqref{eq:flat-gauge} for both cases.

The particle mass $M$ is actually a function on $T^2$ and is given by
\beq
M = \sqrt{D} = \sqrt{D_0} + \frac{D' }{2\sqrt{D_0}} + \cO(D'^2).
\eeq
Thus the real part $\Re S_{\rm particle} $ gives
\beq
\exp\left(- \int_{\sX_y}  M \d s  \right) &= \exp(-\sqrt{D_0} L_1) \left( 1 -  \int_{\sX_y} \d x \frac{D' }{2\sqrt{D_0}} +\cO(D'^2) \right), \nonumber \\
\exp\left(- \int_{\sY_x}  M \d s  \right) &= \exp(-\sqrt{D_0} L_2) \left( 1 -  \int_{\sY_x} \d x \frac{D' }{2\sqrt{D_0}} +\cO(D'^2) \right).
\eeq

The particle path integral contains, among other things, integrals over $y$ for the case $\sX_y$ and $x$ for the case $\sY_x$. 
Therefore, the path integrals in the topological classes of $\sX_y$ and $\sY_x$ contain terms proportional to
\beq
\sX_y~:~& e^{-\sqrt{D_0} L_1 + 2\pi \i  \alpha}  \int_{T^2} \d x \d y\, e^{-2\pi \i m y /L_2} \left( \i  A'_x -\frac{D' }{2\sqrt{D_0}}  \right) , \nonumber \\
\sY_x~:~& e^{-\sqrt{D_0} L_2 + 2\pi \i  \beta}  \int_{T^2} \d x \d y\, e^{2\pi \i m x /L_1} \left( \i  A'_y -\frac{D' }{2\sqrt{D_0}}  \right) .
\eeq
The effective action $\bar S_{\text{eff}}[A,D] $ contains terms proportional to them. 
Because of the factors $e^{-2\pi \i m y /L_2}$ and $e^{2\pi \i m x /L_1} $, these integrals on $T^2$ indeed pick up nonzero momentum Fourier modes in $A'$ and $D'$. 

More explicitly, let us denote the Fourier mode decomposition of $D'$ as
\beq
D' = \frac{1}{L_1L_2}\sum_{(\ell_1, \ell_2) \in \bZ^2 } D_{(\ell_1, \ell_2)} \exp\left[ 2\pi \i \left( \frac{\ell_1 x}{L_1} + \frac{\ell_2 y}{L_2}  \right) \right].
\eeq
The $A'$ is decomposed in the same way.
We can also consider more general topological classes of worldlines of the form 
\beq
\{(k_1 L_1 \tau, k_2 L_2\tau)~|~\tau \in [0,1]\},
\eeq
where $(k_1,k_2)$ are integers.
Then, the effective action contains terms proportional to 
\beq
D_{(-m k_2, m k_1)} \exp\left[ - \sqrt{D_0(k_1^2 L_1^2+k_2^2 L_2^2)} +2\pi \i (k_1   \alpha+k_2 \beta)  \right],
\eeq
and also similar terms for $A'$.

Momentum conservation does not forbid quadratic and higher order terms in $A'$ and $D'$ as in \eqref{eq:A'D'explansion1} even if $ A^{(m)}$ is absent. For notational simplicity let us only consider $D'$. Then the effective action contains terms of the schematic form
\begin{multline}
\bar S_{\rm eff} \supset  c D_{(-m k_2, m k_1)} \exp\left( - \sqrt{D_0} L +2\pi \i \cA  \right)+  \bar c D_{(m k_2, -m k_1)} \exp\left( - \sqrt{D_0} L -2\pi \i \cA  \right)  \\
+  d D_{(-m k_2, m k_1)} D_{(m k_2, -m k_1)}   + \cdots
\end{multline}
where $L=\sqrt{(k_1^2 L_1^2+k_2^2 L_2^2)}$, $\cA=k_1  \alpha+k_2  \beta$, and $c$, $\bar c$ and $d$ are coefficients which are expected to be not exponentially suppressed. Notice that $(k_1,k_2)$ and $(-k_1,-k_2)$ give opposite momentum modes.\footnote{One can also interpret the particle worldline with $(-k_1, -k_2)$ as the anti-particle worldline with $(k_1, k_2)$.}
The saddle point is then given by
\beq
D_{(-m k_2, m k_1)} \sim -\frac{\bar c}{d}  \exp\left( - \sqrt{D_0} L -2\pi \i \cA  \right), \qquad \bar S_{\rm eff} \supset -\frac{c\bar c}{d} \exp\left( - 2\sqrt{D_0} L   \right).
\eeq
As far as $\Re \sqrt{D_0} >0$, they are exponentially suppressed for large $L_1$ and $L_2$.

\subsection{Schwinger effect}\label{app:schwinger}
Here we give another derivation of the exponent of the vacuum decay rate \eqref{eq:schwinger} by using the particle picture. It is computed by using the standard thin wall approximation (see e.g. \cite{Weinberg:1996kr} for a textbook account). In our case, the thin wall is given by a worldline of the charged particle. The derivation here has the advantage of giving an intuitive picture that the vacuum decay rate is indeed due to pair creations of charged particles, but the result is not as precise as the one given in Section~\ref{sec:small}. 
In the context of the $\CPN$ model, the same computation has been done in \cite{Lawrence:2012ua}.

We work in Euclidean space $\bR^2$ (which may be regarded as a large volume limit of $T^2$). Then we consider the worldline $\sC$ of a particle surrounding a bounded region $\sD$. If the electric field $E_{\rm L} = \i E$ is positive, we take the orientation of $\sC$ to be clockwise, i.e., $\partial \sD =  -\sC$. It can also be regarded as the worldline of an anti-particle going in the counterclockwise direction $-\sC$. The action of the particle is given by
\beq
-S_{\rm particle} &= -  \int_\sC  M\d s + \i \int_\sC A \nonumber \\
& = -  \int_\sC  M\d s - \i \int_\sD F,
\eeq
where $F=\d A$ is the field strength 2-form. 

The fact that the particle worldline separates two different (metastable) vacua can be seen as follows. Suppose that the region outside of $\sD$ is in the $n$-th vacuum. 
Recall that the $\theta$-angle appears in the combination $\theta +2\pi n$ as shown in \eqref{eq:afterPoisson}.
On the other hand, the term $- \i \int_\sD F$ in $-S_{\rm particle} $ means that the $\theta$-angle inside the region $\sD$ is effectively shifted by $-2\pi$,
\beq
\i \frac{(\theta+2\pi n)}{2\pi} \int F - \i \int_\sD F = \i \frac{(\theta+2\pi n -2\pi)}{2\pi} \int_{\sD} F +  \i \frac{(\theta+2\pi n)}{2\pi} \int_{\bR^2 \setminus \sD} F.
\eeq
Therefore, the region inside $\sD$ is in the $(n-1)$-th vacuum. Thus the worldline $\sC$ seperates the $n$-th vacuum and the $(n-1)$-th vacuum. 

Let us compute the particle action in the thin wall approximation. As usual, we assume that the bubble solution is spherically symmetric so that the $\sC$ is $S^1$ with radius $r$ and the $\sD$ is the disk surrounded by $S^1$. Then, we get
\beq
-S_{\rm particle}  = - 2\pi r M +\i \pi r^2 E = - 2\pi r \sqrt{D} +  \pi r^2 E_{\rm L},
\eeq
where we have used $F_{xy} =- E = \i E_{\rm L}$ and $M=\sqrt{D}$. Extremizing this action, we get
\beq
r = \frac{ \sqrt{D} }{E_{\rm L}}, \qquad -S_{\rm particle} = - \frac{\pi D}{E_{\rm L}}.
\eeq
The vacuum decay rate is proportional to $\exp(-S_{\rm particle})$.
This action $-S_{\rm particle} = -\pi D/E_{\rm E}$ is precisely the exponent of \eqref{eq:schwinger} for the case $E_{\rm L}>0$. When $E_{\rm L}<0$, we instead consider counterclockwise orientation of $\sC$.

If we look at the worldline $\sC$ from the point of view of time evolution in $\bR^2$, it represents a pair creation of charged particles.\footnote{In Euclidean signature, the pair of charged particles is annihilated after the creation. In Lorentz signature, the charged particles go away from each other.} Thus it is indeed the Schwinger effect.

\section{A formula for the gamma function}\label{app:formula}
In this appendix we show the formula
\beq
  \int_0^{\infty}\left( \frac{1}{2\sinh t} - \frac{1}{2t} \right)\frac{e^{- 2z t}}{t}\d t = \log \Gamma(z+1/2) -z \log z + z- \frac{1}{2}\log2\pi. \label{eq:a-formula}
\eeq  
Our starting point is the Binet's first integral formula for the gamma function 
    \begin{align}
        \log \Gamma(z)=\left(z-\frac{1}{2}\right)\log z -z + \frac{1}{2}\log2\pi + \int_0^{\infty}\left(\frac{1}{2}-\frac{1}{t}+\frac{1}{e^t-1}\right)\frac{e^{-zt}}{t}\d t \, .
        \label{eq:Binet}
    \end{align}
which hold for $\text{Re}[z]>0$~(See Secction 12.3 in \citep{Whittaker_Watson_2021}).
Note that 
    \begin{align}
        &\int_0^{\infty}\left(\frac{1}{2}-\frac{1}{t}+\frac{1}{e^t-1}\right)\frac{e^{-zt}}{t}\d t  \nonumber \\
        &=  \int_0^{\infty}\left( \frac{1}{2\sinh (t/2)} - \frac{1}{t} \right)\frac{e^{-(z+1/2) t}}{t}\d t +  \int_0^{\infty} \left(\frac{e^{- t/2}}{t} -\frac{1}{t}+\frac{1}{2} \right)\frac{e^{-z t}}{t}\d t .\label{eq:calculation10}
    \end{align}
Let us define a function $h(z)$ by
\beq
h(z) =  \int_0^{\infty} \left(\frac{e^{- t/2}}{t} -\frac{1}{t}+\frac{1}{2} \right)\frac{e^{-z t}}{t}\d t .
\eeq    
Its second derivative is easily calculated as
\beq
\frac{\d^2}{\d z^2} h(z) = \int_0^{\infty} \left( e^{- t/2} - 1+\frac{1}{2} t \right) e^{-z t} \d t = \frac{1}{z+1/2} - \frac{1}{z} +\frac{1}{2z^2}.
\eeq    
Therefore,
\beq
h(z) = (z+1/2)\log(z+1/2) - z\log z -\frac12 \log z + c'_0 +c'_1 z,
\eeq
where $c'_0$ and $c'_1$ are constants. These constants are determined by comparing the behavior in the limit $z \to \infty$. The function $h(z)$ goes to zero in the limit $z \to \infty$, and hence
\beq
h(z) = (z+1/2)\log\left(1+\frac{1}{2z} \right)  - \frac12.
\eeq
By using it as well as $\Gamma(z+1) = z\Gamma(z)$, the formula \eqref{eq:Binet} gives
\beq
 & \int_0^{\infty}\left( \frac{1}{2\sinh (t/2)} - \frac{1}{t} \right)\frac{e^{- (z+1/2) t}}{t}\d t  \nonumber \\
&  = \log \Gamma(z+1) -(z+1/2) \log (z+1/2) + (z+1/2)- \frac{1}{2}\log2\pi.
\eeq    
This implies the desired formula \eqref{eq:a-formula}.

\bibliographystyle{ytphys}
\bibliography{main}

\end{document}